\documentclass[preprint]{revtex4}

\usepackage{amsfonts}
\usepackage{amsmath}
\usepackage{amssymb}
\usepackage{graphicx}
\usepackage{psfrag}
\usepackage{epsfig}
\usepackage{xcolor}
\usepackage{relsize}
\usepackage{bm}
\usepackage[normalem]{ulem}

\newcommand{\be}{\begin{equation}}
\newcommand{\ee}{\end{equation}} 
\newcommand{\bea}{\begin{eqnarray}}
\newcommand{\eea}{\end{eqnarray}}

\newcommand{\bsig}{{{\boldsymbol\sigma}}}

\newcommand{\di}{\mathop{}\!\mathrm{d}}
\newcommand{\de}{\mathop{}\!\partial}

\newcommand{\dede}[2]{\frac{\de #1}{\de #2}}

\begin{document}

\title{Theoretical and numerical investigations of inverse patchy
  colloids in the fluid phase}

\author{Yurij V. Kalyuzhnyi}
\email[Author to whom correspondence should be addressed. 
Electronic address: ]{yukal@icmp.lviv.ua}
\affiliation{Institute for Condensed Matter Physics, Ukrainian Academy of 
Science, Svientsitskoho 1, UA-79011 Lviv, Ukraine}
\author{Emanuela Bianchi}
\email{emanuela.bianchi@tuwien.ac.at}
\affiliation{Institute f\"ur Theoretische Physik, Technische Universit\"at 
Wien, Wiedner Hauptstra{\ss}e 8-10, A-1040 Wien, Austria}
\author{Silvano Ferrari}
\email{silvano.ferrari@tuwien.ac.at}
\affiliation{Institute f\"ur Theoretische Physik, Technische Universit\"at
Wien, Wiedner Hauptstra{\ss}e 8-10, A-1040 Wien, Austria}
\author{Gerhard Kahl}
\email{gerhard.kahl@tuwien.ac.at}
\affiliation{Institute f\"ur Theoretische Physik and Center for 
Computational Materials Science (CMS), Technische Universit\"at Wien, 
Wiedner Hauptstra{\ss}e 8-10, A-1040 Wien, Austria}
\date{\today}

\begin{abstract} 

We investigate the structural and thermodynamic properties of a new class of patchy colloids, referred to as inverse patchy colloids (IPCs) in their fluid phase via both theoretical methods and simulations.
IPCs are nano- or micro- meter sized particles with differently charged surface regions.
We extend conventional integral equation schemes to this particular class of systems: our approach is based
on the so-called multi-density Ornstein-Zernike equation, supplemented by the associative Percus-Yevick approximation (APY). To validate the accuracy of our framework, we compare the obtained results with data extracted from $NpT$ and $NVT$ Monte Carlo simulations. 
In addition, other theoretical approaches are used to calculate the properties of the system: the reference hypernetted-chain (RHNC) method and the Barker-Henderson thermodynamic perturbation theory. Both APY and RHNC frameworks provide accurate predictions for the pair distribution functions: APY results are in slightly better agreement with MC data, in particular at lower temperatures where the RHNC solution does not converge.
  
\end{abstract}

\maketitle

\section{Introduction}
\label{sec:introduction}

Inverse patchy colloids (IPCs) have been introduced~\cite{bianchi2011} as a new class of particles within the wide field of colloids with patterned surfaces that are usually referred to as patchy particles~\cite{ilona:review,Bianchi2011r}; IPCs can indeed be seen as patchy particles with charged patches. The IPC model put forward in Ref.~\cite{bianchi2011} originated from the idea that positively charged star polymers adsorbed on the surface of negatively charged colloids give rise to complex units with differently charged surface regions~\cite{Blaak_2008}; the coarse-grained model developed for such systems is nonetheless generally valid to describe colloids with a heterogeneous surface charge~\cite{peter}. 

IPCs are characterized by highly complex spatial and orientational pair interactions~\cite{bianchi2011} due to the fact that regions of unlike charges attract each other, while regions of like charges mutually repel. The calculation of the effective interactions between IPCs -- evaluated via simple electrostatic considerations -- leads to expressions in terms of truncated series expansions that are not amenable to investigations in extended ensembles via either computer simulations or theoretical approaches. However, appropriate coarse-graining schemes have been proposed such that the effective interactions reduce to simple analytic expressions, which nevertheless include the characteristic features of the original model~\cite{bianchi2011}, thus allowing many body simulation approaches and theoretical investigations. 

The intricate shape of the interaction potential (which can be tuned via either the decoration of the particles or the properties of the solvent) is responsible for the self-assembly of IPCs into well-defined structures at mesoscopic length scales. In striking contrast to conventional patchy particles, the assembly behavior of IPCs can be selectively addressed by easily accessible, external parameters: in recent studies on IPCs systems confined to planar quasi two-dimensional geometries, it was shown that the self-assembly scenarios of IPCs can be reversibly tuned by e.g. minute pH changes of the solution~\cite{bianchi:2d2013,bianchi:2d2014}.  Further, several investigations in the bulk have shown that IPCs can form -- apart from different types of disordered phases (such as gas, liquids, or gels) -- a broad variety of highly complex ordered phases~\cite{Guenther_thesis_2012,ismene,silvano30n,eva,silvano45n}. An exact localization of the phase boundaries of competing phases is extremely difficult and expensive from both the methodological and the computational point of view, as demonstrated in the comprehensive evaluation of the thermodynamic properties of a particular system of IPCs forming a lamellar bulk phase~\cite{ismene,eva}.

In view of the complexity of IPCs systems, it is advisable to search for methods that are computationally cheaper than simulations. Our theoretical approach is based on the ideas proposed by Wertheim in the 80s: to describe the properties of associating fluids, Wertheim used a model system with an isotropic steric repulsion complemented with patch-patch attractive interactions~\cite{wertheim1,wertheim2,wertheim1986a,wertheim1986b}. In a successive step, thanks to the modelization of patch-center interactions introduced by Nezbeda~\cite{nezbeda}, Wertheim further extended his theory~\cite{wertheim1986dim}. Wertheim's concept for associating liquids is ideally suited to describe the structural and thermodynamic properties of patchy particles and forms the foundation on which our work is based.
Our Wertheim-type approach relies on an analogue of the Ornstein-Zernike (OZ) equation~\cite{HeM} expanded in terms of so-called bonded correlation functions, indexed by the number of bonds that a particle actually forms; such an expansion is referred to as the multi-density OZ equation. The original associative scheme is very versatile and can be applied to any type of patch decoration; the complexity is drastically reduced when particles are decorated only by a few, equivalent patches, as it is the case in our IPC systems. By introducing in addition the ideal network approximation and an extension of the Percus-Yevick or the hypernetted-chain closure, we outline a computationally cheap iterative scheme, referred to in the following as APY or AHNC, respectively.

Our associative description is compared to two other well known and widely used theories, namely the reference hypernetted-chain approach (RHNC)~\cite{Giacometti} and the Barker-Henderson thermodynamic perturbation theory (BH-TPT)~\cite{Gogelain}. Similar to our associative theory, also RHNC relies on the molecular OZ equation but in this case the expansion of the correlation functions occurs in terms of rotational invariants; in contrast, BH-TPT is a simple perturbative description of the free energy of the system.

In the present manuscript we have two main focuses. On one side, we investigate to what degree of accuracy the APY approach is able to describe the features of two selected IPC models in the fluid phase. On the other side, we study how the choice of the mapping procedure, used to derive the coarse-grained potential parameters from the analytical effective interaction, affects the static properties of IPC systems. As shown in Ref.~\cite{bianchi2011}, the same effective interaction can be coarse grained in different ways, thus affecting the value of the contact energies, the patch size and the particle interaction range. We focus here on two different mappings of a microscopic IPC system presented in Ref.~\cite{bianchi2011} and we investigate its structural and thermodynamic properties. We perform Monte Carlo (MC) simulations which provide reference data. Both the APY and RHNC frameworks are able to give a faithful description of the simulation results, APY being slightly more accurate and having a broader convergence range. The main difference between the two chosen systems relies in the patch bonding volume: when this volume is relatively small, APY works better since the theory neglects multiple bonds per patch; nonetheless, also for IPCs with a bigger bonding volume, the associative theory provides a good description of the static observables. 

The paper is organized as follows. In Sec.~II the general associative description is presented together with the expressions for the thermodynamic properties of the system in terms of the distribution functions; we consider here the general case of IPCs with $n_s$ equivalent patches. In Sec.~III we apply the developed formalism to study the two-patch version of the model. Here we specify the potential model parameters for the coarse-grained version of the IPC model and report details of the MC simulation method, RHNC and BH-TPT theories. In Sec.~IV we present our results and conclusions are collected in Sec.~V.

\section{The Theory for the general model}
\label{sec:theory}

\subsection{The general model}
\label{subsub:general_model}

In the following we present a general model for an IPC, which consists of a spherical particle decorated by an
arbitrary number ($n_s$) of patches, called off-center interaction
sites. The pair potential between two interacting IPCs is given by
$U(1, 2)$, where $1, 2 = \{ {\bf r}_{1,2}, \boldsymbol{\omega}_{1,2} \}$ denotes
the spatial as well as the orientational degrees of freedom of
particle 1, 2. $U(1, 2)$ consists of a spherically symmetric
potential, $U_{00}(r)$, where $r$ is the distance between particle 1 and 2, 
acting between the centers of the particles and an orientationally dependent potential
due to the $n_s$ off-center interaction sites:

\begin{equation}
U(1, 2) = U_{00}(r) + \sum_K \Big[ U_{K0}(1, 2) +
U_{0K}(1, 2) \Big] + \sum_{KL} U_{KL}(1, 2) .
\label{eq:utot}
\end{equation}

Here the index $0$ denotes the particle centers and capital letters
(such as $K$ or $L$) specify off-center sites. The set of all these
sites will be denoted as $\Gamma$ and subsets of $\Gamma$ will by
specified by small Greek letters, i.e. $\alpha, \beta,$ etc.  It is
assumed that the center-site potential, $U_{0K}(1, 2)$, is attractive
and short-ranged, such that each site $K$ of one particle can be
bonded only to one center of another particle 
and two sites of one particle cannot be simultaneously bonded to the 
center of the other particle. Furthermore, it is assumed that
the site-site potential, $U_{KL}(1, 2)$, is repulsive and short-ranged.

Our particular choice of potentials (i.e., their particular functional
form, the positions of the attractive sites, etc.)  will be discussed
in Section III, where our theory will be applied to describe the
properties of a specific IPC model, recently proposed in
Ref.~\cite{bianchi2011}.

We consider $N$ particles, confined in a volume $V$, at temperature
$T$, and pressure $p$; the homogeneous number density is $\rho = N/V$.

\subsection{Diagrammatic analysis and topological reduction of the 
general model}
\label{subsec:diagram_topological}

In the diagrammatic analysis of the general model introduced above we
will follow a scheme developed earlier to describe models with
multiple site-site bonding~\cite{wertheim1986a,wertheim1986b}
and combine it with a framework put forward for a model
with site-center bonding~\cite{wertheim1987, kalyuzhnyi1991}.

For the sake of our theoretical analysis we split up the total pair
potential, $U(1, 2)$, into a reference and an associative part, i.e.,
$U_{\rm ref}(1, 2)$ and $U_{\rm ass}(1, 2)$, respectively

\begin{equation}
U(1, 2) = U_{\rm ref}(1, 2) + U_{\rm ass}(1, 2),
\label{eq:utot}
\end{equation}
where

\begin{equation}
U_{\rm ref}(1, 2) = U_{00}(r) + \sum_{KL} U_{KL}(1, 2)
\label{eq:ref}
\end{equation}
and

\begin{equation}
U_{\rm ass}(1, 2) = \sum_K \Big[ U_{K0}(1, 2) + U_{0K}(1, 2) \Big].
\label{eq:ass}
\end{equation}

Consequently, the Mayer function (or Mayer $f$-bond), $f(1, 2) = \exp
\left[ - \beta U(1, 2) \right] - 1$, can be decomposed as follows

\begin{eqnarray}
\label{eq:ftot}
f(1, 2) & = & f_{\rm ref}(1, 2) +e_{\rm ref}(1, 2)
\left \{ \prod_{K} \left[ f_{K0}(1, 2) + 1 \right]
\prod_L \left[ f_{0L}(1, 2) + 1 \right] - 1 \right\} \\ \nonumber
& = & f_{\rm ref}(1, 2) + \sum_K F_{K0}(1, 2) +
\sum_L F_{0L}(1, 2) + \sum_{KL} F_{KL}(1, 2) 
\end{eqnarray}
where

\begin{equation}
e_{\rm ref}(1, 2) = \exp \left[ - \beta U_{\rm ref}(1, 2) \right] ~~~~~ 
{\rm and} ~~~~~
f_{\rm ref}(1, 2) = e_{\rm ref}(1, 2) - 1 ,
\end{equation} 

\begin{equation}
f_{K0}(1, 2) = \exp \left[ - \beta U_{K0}(1, 2) \right] - 1,
\end{equation}

\begin{equation}
F_{K0}(1, 2) = e_{\rm ref}(1, 2) f_{K0}(1, 2) ~~~~
F_{KL}(1, 2) = e_{\rm ref}(1, 2) f_{K0}(1, 2) f_{0L}(1, 2).
\label{eq:F}
\end{equation}

We now introduce the activity $z$ via

$$ 
z = \Lambda^{-3} {\rm e}^{\beta \mu}
$$
where $\Lambda$ is the de Broglie wavelength, $\beta = 1/(k_{\rm B}T)$
(with $k_{\rm B}$ being the Boltzmann constant) and $\mu$ is the
chemical potential. 
Note, that in the decomposition (\ref{eq:ftot}) we neglect terms 
containing the products $f_{K0}(12)f_{L0}(12)$ or/and $f_{0K}(12)f_{0L}(12)$, which
describe the bonding of the two patches of one particle to the center of the other.
In the diagrammatic expansion of the grand
partition function of the system, $\Xi$, in terms of the activity $z$,
each Mayer $f$-bond will be substituted by either $f_{\rm ref}$-bonds
or by products of $e_{\rm ref}(1, 2)$ with one or more $f_{K0}$-
or/and $f_{0K}$-bonds. We assume that each of the sites can be bonded only once.

Usually in diagrammatic expansions the particles are depicted by a
circle; however for our problem it is more convenient to introduce in the
diagrammatic expressions {\it hypercircles} (which now represent the
particles) instead of circles~\cite{wertheim1986a}: each hypercircle
is depicted as an open circle that contains small circles, denoting
the sites.  Now the cluster integrals that enter the diagrammatic
expansion of $\Xi$ are represented by the collection of field ${\tilde
  z}$-hypercircles, connected by $f_{\rm ref}$- and $e_{\rm
  ref}$-bonds in parallel to one or more $f_{K0}$- and/or
$f_{0K}$-bonds. Here

$$
\tilde z (i) = z {\rm e}^{- \beta U(i)}
$$
denotes the spatially and orientationally dependent activity and
$U(i)$ is a possible external field acting on particle $i$;
again, this index stands for the spatial and the orientational
degrees of freedom of this particle. For a uniform system $\tilde z(i)
= z$.

\begin{figure}[h] 
\centering 
\includegraphics[width=\columnwidth, clip=true]{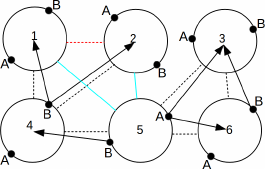}
\caption{(Color online) Example of a typical diagram as it emerges in the diagrammatic expansion of $\ln \Xi$: it corresponds to an hexamer, where each particle is represented by a hypercircle. The diagram is depicted for a general two-patch model where the small full circles denoted by A and B represent the patches; $f_{0K}$- and/or $f_{K0}$-bonds are denoted by solid arrows pointing from site (patch) $K$ of one hypercircle to the center of another hypercircle, $e_{\rm ref}$-bonds by dashed lines and $f_{\rm ref}$-bonds by solid lines.  Diagrams are built according to a three-step procedure; here step (i) is drawn in black, step (ii) in red, and step (iii) in cyan. Step (i) corresponds to drawing all possible combinations of $f_{0K}$- and/or $f_{K0}$-bonds, and to subsequently adding an $e_{\rm ref}$-bond between the particles that have been connected. Step (ii) corresponds to adding an $e_{\rm ref}$-bond to all the pairs of particles bonded to the same patch of a third particle, unless they are already linked: in the depicted example particles 1 and 2 are both connected to patch B of particle 4, so an $e_{\rm ref}$-bond is added; this procedure does not apply to particles 3 and 6 because they have been already linked by an $e_{\rm ref}$-bond in step (i). In step (iii) we consider all possible combinations of $f_{\rm ref}$-bonds between pair of particles that have not already been directly linked by any kind of bond. 
}
\label{fig:diagram}
\end{figure}

In Figure~\ref{fig:diagram} we show a diagram
representing an ensemble of $s$ hypercircles, referred to as $s$-mer.  
In the figure different types of bonds, i.e.  $f_{\rm ref}$-, $e_{\rm ref}$-, $f_{0K}$- and $f_{K0}$-bonds, are distinguished by different lines. 
Diagrams are constructed
via the three following steps: (i) generate the subset of all possible
connected diagrams with $f_{K0}$- and $f_{0K}$-bonds and insert an
$e_{\rm ref}$-bond between hypercircles directly connected by
either a $f_{K0}$- or a $f_{0K}$-bond; (ii) insert an 
$e_{\rm ref}$-bond between all pairs of not directly connected hypercircles, whose centers are connected via either a $f_{K0}$- or a $f_{0K}$-bond to the same site $K$ of a third particle; (iii) consider
all possible ways of inserting $f_{\rm ref}$-bonds between the pairs
of hypercircles which have not been not directly connected during the previous two steps.

As a result the diagrams, which appear in the expansions for
$\ln{\Xi}$ and for the one-point density $\rho(1)$, defined as
\begin{equation}
\rho(1) = \tilde z(1)
\frac{\delta \ln \Xi}{\delta \tilde z(1)}
\end{equation}
can be expressed in terms of $s$-mer diagrams, as follows

\begin{eqnarray}
\ln\Xi & = & ~sum~of~all~topologically~distinct~connected~diagrams~
  consisting~of~\\ \nonumber
& & s-mer~diagrams~with~s=1,\ldots,\infty~and~f_{\rm
ref}-bonds~between~pairs~of~\\ \nonumber
& & hypercircles~in~distinct~s-mer~diagrams; \nonumber
\label{eq:statement_A}
\end{eqnarray}

\begin{eqnarray}
\rho(1) & = & sum~of~all~topologically~distinct~connected~diagrams~
\\ \nonumber
& & obtained~from~\ln \Xi~by~replacing~in~all~possible~ways~one~field~
  {\tilde z}-hypercircle \\ \nonumber
& &  by~a~{\tilde z}(1)-hypercircle~labeled~1. \nonumber
\label{eq:statement_B}
\end{eqnarray}

The diagrams appearing in the expression for the one-point density
$\rho(1)$ can be classified with respect to the set of bonded sites at
the labeled point 1. We denote the sum of the diagrams with the set of
the bonded sites $\alpha$ at the labeled hypervertex as
$\rho_\alpha(1)$. Thus we have

\begin{equation}
\rho(1) = \sum_{\alpha \subseteq \Gamma} \rho_\alpha(1) .
\label{eq:rotot}
\end{equation}

Here the terms with $\alpha = 0$ (where 0 identifies the subset of diagrams where sites have no bonds) and $\alpha=\Gamma$ (where $\Gamma$ is the subset of diagrams where all sites are bonded) are included. 
Note that the term
$\rho_0(1)$ denotes the one-point density of particles without bonded
sites at 1; however, this does not mean that particle 1 is unbonded, since
its center may still be bonded to any number of sites belonging to
other particles.

Now we will apply the procedure of topological reduction to switch
from an expansion in terms of ${\tilde z}(i)$ to a multi-density
expansion. Following Wertheim's work~\cite{wertheim1986a} we introduce
operators (which are denoted by $\varepsilon$'s) as
follows: we associate with each labeled hypercircle $i$ an operator
$\varepsilon_\alpha(i)$ with the properties

$$
\varepsilon_\alpha(i) = \prod_{K \in \alpha} \varepsilon_{ K }(i) ~~~~~
\alpha \subseteq \Gamma
$$

\begin{equation}
\varepsilon_{K}^2(i) = 0 ~~~~ \varepsilon_0(i) = 1.
\label{eq:eps}
\end{equation}
for all off-center sites $K$. 

Now any one- or two-point quantity -- denoted by $a(1)$ or $b(1, 2)$, respectively -- can be presented in the operator notation in the form

\begin{equation}
\hat{a}(1) = \sum_{\gamma \subseteq \Gamma} \varepsilon_\gamma(1)a_\gamma(1),
\label{eq:a}
\end{equation}

\begin{equation}
\hat{b}(1, 2) = \sum_{\gamma, \lambda \subseteq \Gamma}
\varepsilon_\gamma(1) b_{\gamma \lambda}(1, 2)\varepsilon_\lambda(2),
\label{eq:b}
\end{equation}
where the hat denotes an expansion of the respective quantity in terms of the operators $\varepsilon_\alpha(i)$.

The usual algebraic rules for linear and bi-linear terms apply to
these expressions; further, analytic functions of these quantities are
defined via their corresponding power series. 

Here we also define an operation that will be useful below

\begin{equation}
a_\Gamma(1) = \langle \hat{a}(1) \rangle_{1}.
\label{eq:brack}
\end{equation}
In this relation, the angular brackets mean that as a consequence of
this operation only the coefficient of the operator
$\varepsilon_\Gamma(1)$ is retained in the expression for $\hat{a}(1)$.

Analyzing the connectivity of the diagrams in $\rho(1)$ at a labeled
hypercircle ${\tilde z}(1)$ we find~\cite{wertheim1986a,wertheim1986b}

\begin{equation}
\hat{\rho}(1)/{\tilde z}(1) = \exp{\left[ \hat{c}(1) \right]} ;
\label{eq:rhoexp}
\end{equation}
$c_\alpha(1)$ (with $\alpha \ne 0$), appearing in the expansion of
$\hat c(1)$, denotes the sum of diagrams in the function
$\rho_{\alpha}(1)/\rho_0(1)$ for which the labeled hypercircle 1 is
not an articulation circle. Similarly, $c_0(1)$ denotes the sum of
diagrams in the function $\rho_0(1)/{\tilde z} (1)$ for which hypercircle
1 is not an articulation circle.

To remove the diagrams containing field articulation circles 
we will follow the earlier studies~\cite{wertheim1986a,wertheim1986b} and
switch from the expansions of $\ln \Xi$ and of $\rho(1)$ in terms of the
activity to density expansions using the following rule: in all
irreducible diagrams appearing in $\hat c(1)$ each field hypercircle
${\tilde z}$, with the bonding state of its sites represented by the
set $\alpha$, is replaced by a $\sigma_{\Gamma - \alpha}(1)$
hypercircle, where the minus sign denotes henceforward the set-theoretic
difference sign of two sets; these new quantities $\sigma_\alpha(1)$ are related to the
densities $\rho_\alpha(1)$ via~\cite{wertheim1986a}.
\begin{equation}
\hat{\sigma}(1) = \hat{\rho}(1) \sum_{\alpha \subseteq \Gamma} \varepsilon_\alpha(1).
\label{eq:sigma}
\end{equation}

This relation can be inverted by formally expanding $\left[
  \sum_{\alpha \subseteq \Gamma} \varepsilon_\alpha(1) \right]^{-1}$
in a power series and by retaining the first term

\begin{equation}
\hat{\rho}(1) = \hat{\sigma}(1)
\prod_{K \in \Gamma} \left[ 1 - \varepsilon_{K}(1) \right].
\label{eq:sigmainv}
\end{equation}

Now the diagrammatic expansions for $c_\alpha(1)$ introduced above can be expressed in terms of irreducible diagrams. To present this result in a compact and convenient form we introduce the fundamental diagrams $c^{(0)}$ defined as follows

\begin{eqnarray}
c^{(0)} & = & sum~of~all~topologically~distinct~irreducible~diagrams 
\\ \nonumber
& &  consisting~of~s-mer~diagrams~with~s=1,\ldots,\infty~and~
  f_{\rm ref}-bonds~ \\ \nonumber
& & between~pairs~of~hypercircles~in~distinct~
  s-mer~diagrams. \\ \nonumber
& & All~hypercircles~are~field~circles~carrying~the~
  \sigma-factor~ \\ \nonumber
& & according~to~the~rule~formulated~above . \nonumber
\label{eq:statement_C}
\end{eqnarray}

The $c_\alpha(1)$ can be obtained by functional differentiation of $c^{(0)}$ with respect to $\sigma_{\Gamma - \alpha}(1)$
\begin{equation}
c_\alpha(1) ={\delta c^{(0)}\over \delta \sigma_{\Gamma - \alpha}(1)}.
\label{eq:fder}
\end{equation}

Now we are able to rewrite the regular one-density virial expansion
for the pressure $p$ and for the Helmholtz free energy $A$ in terms of the
density parameters $\hat{\sigma}(1)$ defined in Equation
(\ref{eq:sigma}). Following a scheme proposed by Wertheim
\cite{wertheim1986a} we obtain
\begin{equation}
\beta pV = \int \left[
\rho(1) - \sum_{\alpha \subseteq \Gamma}
\sigma_{\Gamma - \alpha}(1) c_\alpha(1) \right] \; d 1 + c^{(0)}.
\label{eq:pvir}
\end{equation}
and
\begin{equation}
\beta A = \int \left[
\rho(1) \ln{{\sigma_0(1)\over \Lambda}} - \rho(1) +
\sum_{\alpha \subseteq \Gamma, \alpha \neq 0}
\sigma_{\Gamma - \alpha}(1) c_\alpha(1)\right] \; d 1 - c^{(0)}.
\label{eq:A}
\end{equation}
These expressions satisfy the standard thermodynamic relations

$$
\rho = \beta \left( \frac{\partial p}{\partial \mu} \right)_T
\qquad \qquad
\frac{A}{V} = \rho \mu - p ,
$$
where the homogeneous density $\rho$ is recovered via

$$
\rho = \int \rho(1) d1 .
$$ 

\subsection{Integral equation theory for the general model}
\label{subsec:ie}

\subsubsection{Multi-density Ornstein-Zernike equation}
\label{subsubsec:MDOZ}

So far our analysis was focused on one-point quantities. Now we
proceed to the corresponding analysis of two-point quantities. In
particular we consider the pair correlation function $h(1, 2)$ which
can be calculated via the following functional derivative~\cite{HeM}

\begin{equation}
\rho(1) h(1, 2) \rho(2) = 
{\tilde z}(1) {\tilde z}(2)
{\delta^2 \ln{\Xi} \over \delta{\tilde z}(1) \delta{\tilde z}(2)};
\label{eq:h}
\end{equation}
here the diagrammatic expansion for $\ln{\Xi}$ -- introduced in the
preceding subsection -- Equation (10)
-- has to be used. Elimination of the diagrams containing articulation circles can
be realized following the topological reduction scheme described above.
Via this route we obtain for the final expression for the pair
correlation function in operator notation 

\begin{equation}
\rho(1) h(1, 2) \rho(2) = \langle {\hat\sigma}(1)
\hat{h}(1, 2) {\hat\sigma}(2) \rangle_{1, 2} .
\label{eq:h_oper}
\end{equation}
In the case of two particle quantities the subscripts on the brackets
denote the arguments to which the procedure specified in Equation
(\ref{eq:brack}) has to be applied.

Alternatively, in the regular notation one finds

\begin{equation}
\rho(1) h(1, 2) \rho(2) = \sum_{\alpha, \beta \subseteq \Gamma} \sigma_{\Gamma -
  \alpha}(1) h_{\alpha \beta}(1, 2) \sigma_{\Gamma - \beta}(2) .
\label{eq:h_usual}
\end{equation}
Here the partial correlation functions, $h_{\alpha\beta}(1, 2)$,
represent those diagrams that have sets of bonded sites $\alpha$ and
$\beta$, which belong to hypercircles 1 and 2, respectively.

On differentiating the diagrammatic expansion for $c^{(0)}$, defined
in statement (20
), we obtain the analogue of the partial direct pair correlation functions~\cite{HeM}

\begin{equation}
c_{\alpha\beta}(1, 2) = 
{\delta^2c^{(0)} \over \delta\sigma_{\Gamma - \alpha}(1)
\delta\sigma_{\Gamma - \beta}(2)} ~~~~~ \alpha, \beta \subseteq \Gamma .
\label{c}
\end{equation}
The functions satisfy an Ornstein-Zernike (OZ) like
integral equation, which -- using the operator notation -- 
can be written as~\cite{wertheim1986a,wertheim1986b}

\begin{equation}
\hat{h}(1, 2) = \hat {c}(1, 2) +
\int \langle \hat c(1, 3) \hat \sigma(3) \hat h(3, 2) \rangle_3 d3 .
\label{OZoper}
\end{equation}

In addition to the partial direct and total correlation functions,
this equation involves also the set of the density parameters
$\sigma_\alpha(1)$ introduced in Equation (\ref{eq:sigma}) which are
not known in advance. The self-consistent relation between the
densities follows from Equations (\ref{eq:rhoexp}) and
(\ref{eq:sigmainv}) with $c_\alpha(1)$ expressed in terms of the pair
correlations. We find for the regular notation~\cite{wertheim1986a,wertheim1986b}
%
%

\begin{equation}
c_\alpha(1) = \sum_{\gamma \subseteq \Gamma} \int g_{\alpha - A, \gamma}(1)
f_{A0}(1, 2) \sigma_{\Gamma - \gamma}(2) d2 .
\label{cusual}
\end{equation}
Here we have introduced the partial distribution functions
$g_{\alpha\beta}(1, 2)$, defined as

\begin{equation}
g_{\alpha\beta}(1, 2) = h_{\alpha\beta}(1, 2) + 
\delta_{\alpha, 0} \delta_{\beta, 0} .
\label{g}
\end{equation}

To close the system of equations, we require an additional link between the
partial direct and total correlation functions, known in literature as
a closure relation: it emerges from the diagrammatic analysis of the
cavity correlation functions, $y_{\alpha\beta}(1, 2)$, defined in
operator notation~\cite{wertheim1986a,wertheim1986b} as

\begin{equation}
\hat{y}(1, 2) = \exp{\left[\hat{t}(1, 2)\right]} ;
\label{eq:t}
\end{equation}
here $\hat{t}(1, 2)$ is the sum of all topologically distinct,
irreducible diagrams consisting of two so-called white hypercircles
labeled 1 and 2, corresponding to coordinates that are not integrated
over. Such a formulation is possible since $\hat t(1, 2)$ does not
have white articulation pairs and no two white circles are adjacent~\cite{HeM,wertheim1986b}.

The relation between the partial cavity correlation functions and the
partial pair distribution functions is established via the following
equation~\cite{wertheim1986a,wertheim1986b}

\begin{equation}
\hat g(1, 2) = e_{\rm ref}(1, 2) \hat y(1, 2) \exp \left[ \hat f(1, 2)
\right]
\label{y}
\end{equation}
with

\begin{equation}
\hat f (1, 2) = \sum_{K \in \Gamma}
\left[ \varepsilon_K(1) \varepsilon_{10}(1, 2) f_{K0}(1, 2) +
f_{0K}(1, 2) \varepsilon_{01}(1, 2) \varepsilon_{K}(2) \right] .
\label{f}
\end{equation}
Here we have introduced the two-point operator $\epsilon_{10}(1, 2)$
in order to prevent bonding of several sites of the same particle
to the center of the other particle. This operator has the following
properties

\begin{equation}
\label{UHC}
\varepsilon_{10}(1, 2)\varepsilon_{01}(1, 2)=1 ~~~~~
\varepsilon_{10}^2(1, 2)=\varepsilon_{01}^2(1, 2)=0 .
\end{equation}

Extracting the subset $\hat{E}(1, 2)$ from the set of diagrams
$\hat{t}(1, 2)$ which have no nodal circles leads to the following
expression for the cavity correlation function

\begin{equation}
\hat{y}(1, 2) = \exp{\left[\hat{N}(1, 2) + \hat{E}(1, 2)\right]} ;
\label{eq:yNE}
\end{equation}
where $\hat{N}(1, 2)$ is the subset of diagrams with nodal circles and
is equal to the convolution term in the OZ equation (\ref{OZoper})

\begin{equation}
\hat{N}(1, 2)=\hat{h}(1, 2)-\hat{c}(1, 2) .
\label{N}
\end{equation}
Combining Equations (\ref{N}), (\ref{eq:yNE}), and (\ref{y}) we finally
obtain

\begin{equation}
\hat{g}(1, 2) = e_{\rm ref}(1, 2)
\exp{\left[\hat{h}(1, 2)-\hat{c}(1, 2)+\hat{E}(1, 2)+\hat{f}(1, 2)\right]}.
\label{gfinal}
\end{equation}

Once $\hat{E}(1, 2)$ is given, we can derive the analogues of the commonly used closure relations
in standard integral equation theory.

\subsubsection{Associative hypernetted-chain and associative 
Percus-Yevick approximations}

A hypernetted-chain (HNC)-like approximation to Equation
(\ref{gfinal}) can be derived by setting $\hat{E}(1, 2) = 0$, i.e.

\begin{equation} 
\hat{g}(1, 2) = e_{\rm ref}(1, 2)
\exp{\left[\hat{h}(1, 2)-\hat{c}(1, 2)+\hat{f}(1, 2)\right]}.
\label{gHNC}
\end{equation}

On the other hand, if we assume that all possible products of the
irreducible diagrams of $\hat t(1, 2)$ are canceled by the diagrams in
$\hat{E}(1, 2)$, i.e.,

\begin{equation}
\hat E(1, 2) + \exp[\hat t(1, 2)] - 1 - \hat t(1, 2) = 0
\label{cancel}
\end{equation}
we obtain a Percus-Yevick (PY)-like approximation, 

\begin{equation}
\hat{y}(1, 2) = \hat{g}(1, 2) - \hat{c}(1, 2).
\label{PY}
\end{equation}

In what follows we will refer to these closure relations as to the
associative HNC (AHNC) or the associative PY (APY) approximations.

\section{IPC model with two equivalent patches}
\label{sec:model}

In this section we will apply the theory developed above to the
evaluation of the structural and the thermodynamic properties of
a recently proposed two-patch version of the IPC
model \cite{bianchi2011}.  Our theoretical results will be compared
with data obtained in computer simulations as well as with results
originating from the reference hypernetted-chain (RHNC) integral
equation theory \cite{Lado} and from the Barker-Henderson
thermodynamic perturbation theory \cite{Gogelain} (BH-TPT).

\subsection{Potential model}
\label{subsec:potential_model}

Recently, a coarse-grained model for colloids has been put forward,
characterized by an axially symmetric surface charge distribution due
to the presence of two polar patches of the same charge, $Z_{\rm p}$, and an
equatorial region of opposite charge, $Z_{\rm c}$ \cite{bianchi2011}.  The
model takes into account the three different regions on the particle
surface and is characterized by three independent parameters: the
range and the strengths of the interaction -- reflecting the
screening conditions and the ratio $Z_{\rm p}/Z_{\rm c}$, respectively
-- and the patch surface coverage \cite{bianchi2011}. 
The choice of the parameters takes advantage of the analytical description
of the microscopic system that was developed in parallel by
extending the concepts of the Debye-H\"uckel
theory~\cite{debyehueckel}.

\begin{figure}[h] 
\centering
\includegraphics[width=\columnwidth, clip=true]{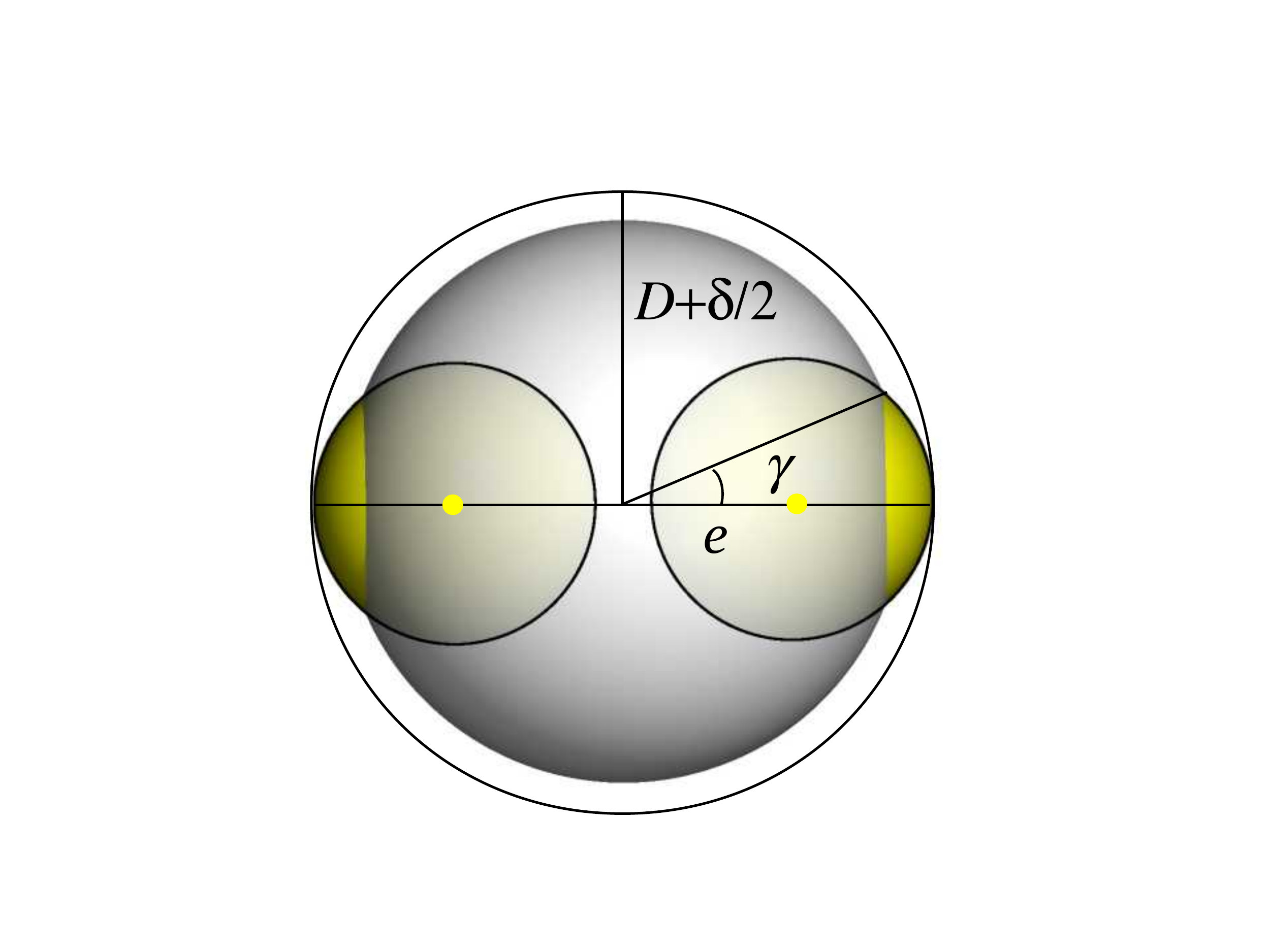}
\caption{(Color online) Schematic representation of a two-particle IPC model.  
  The dark gray sphere features the colloidal particle
  and the small yellow points, located inside the colloid, 
  represent the two interaction sites. The yellow caps correspond to the interaction areas, 
  while the interaction
  sphere of the bare colloid is delimited by the black outermost
  circle. The relevant parameters of the system are the particle diameter,
  $D$, the interaction range, $\delta$, the distance between the
  interaction sites and the colloid center, $e$, and the half opening
  angle, $\gamma$, which defines the patch extension on the particle
  surface.}
\label{fig:IPCmodel}
\end{figure}

The coarse-grained model put forward in Ref. \cite{bianchi2011}
features an IPC as a spherical, impenetrable colloidal particle (of
diameter $D$ and central charge $Z_{\rm c}$) carrying two interaction
sites (each of charge $Z_{\rm p}$) located at distances $e$ ($< D/2$)
in opposite directions from the particle center (see
Figure \ref{fig:IPCmodel}). The two patches are thus positioned at the poles
of the particle, the remaining bare surface of the colloid will be referred to
as the equatorial region. The electrostatic screening
conditions (expressed via the Debye screening length $\kappa^{-1}$)
determine the range $\delta$ of the pair interaction independently of
the relative orientation of the particles. For each parameter set
$(D,e,\delta)$ the patch size, defined by the opening angle $\gamma$,
is uniquely determined by Equations (10) and (11) of
Ref. \cite{bianchi2011}.  The energy strengths are set by mapping
the coarse-grained
potential to the analytical Debye-H\"uckel potential developed for
IPCs in water at room temperature \cite{bianchi2011}. As in Ref. \cite{bianchi2011},
we considered here only overall neutral particles,
i.e. ${\rm Z}_{\rm tot}={\rm Z}_{\rm c}+2{\rm Z}_{\rm p} = 0$. In
contrast to Ref. \cite{bianchi2011}, the coarse-grained pair potential
is further normalized such that the minimum of the equatorial-polar
attraction ($\epsilon_m$) sets the energy unit. The final expression for the pair
potential acting between particles is given by

\begin{eqnarray}
U_{00}(r_{00}) & = & U_{\rm HS}(r_{00}) + 
\frac{4\epsilon_{00}}{ \epsilon_m D^3} \left( 2R_0+{1\over 2}r_{00} \right)
\left( R_0-{1\over 2}r_{00} \right)^2 \Theta(2R_0-r_{00}), \\
U_{01}(r_{01}) & = & \frac{2\epsilon_{01}}{\epsilon_m D^3}
\left\{ \left[ 2R_0+{1\over 2r_{01}}
\left(R_+R_-+r_{01}^2\right) \right]
\left[ R_0-{1\over 2r_{01}} \left( R_+R_-+r_{01}^2 \right) \right]^2 +
\right. \\
& & \left.
+\left[ 2R_1+{1\over 2r_{01}} \left( R_+R_--r_{01}^2 \right) \right]
\left[ R_1-{1\over 2r_{01}} \left( R_+R_--r_{01}^2 \right) \right]^2
\right\} \times \\ \nonumber
& & ~~~~~~~~~ \times  \Theta(R_+-r)\Theta(r-R_-), \\ \nonumber
U_{11}(r_{11}) & = & \frac{4\epsilon_{11}}{\epsilon_m D^3} \left( 2R_1+{1\over 2}r_{11} \right)
\left( R_1-{1\over 2}r_{11} \right)^2 \Theta(2R_1-r_{11}) .
\end{eqnarray}
Here, $r_{00},r_{01}$, and $r_{11}$ are the distances between the
particles centers, between the center and an attractive site, and
between the sites, respectively; $U_{\rm HS}(r_{00})$ is the
hard-sphere potential, $\epsilon_{00},\epsilon_{01},$ and
$\epsilon_{11}$ are the corresponding energy strength parameters,
$\Theta(x)$ is the Heaviside step function, $R_0=(D+\delta)/2$,
$R_1=R_0-e$, $R_+=R_0+R_1$, $R_-=R_0-R_1$; $\delta$ and $e$ have been
defined above (see also Figure \ref{fig:IPCmodel}).

\subsection{Multi-density OZ equation for the IPC model with $n_s$
equivalent patches. \\ Ideal Network Approximation}
\label{subsec:multidensity_OZ}

Expressions obtained in subsection \ref{subsec:ie} are general
and thus applicable to a number of different versions of the
model. Here our goal is to apply this formalism to study the structure
and thermodynamics of the IPC model proposed in
Ref. \cite{bianchi2011}, having $n_s$ equivalent attractive sites.
We note that in absence of double bonding $F_{KL}(1, 2)$ is zero. We assume that double bonding does not occur in the IPC models considered here.

A straightforward application of expression (\ref{eq:b}) and of
the corresponding OZ equation (\ref{OZoper}) to our particular model
will turn the correlation functions into matrices with $2^{n_s}\times
2^{n_s}$ elements; taking into account the equivalence of the sites
(i.e., patches), this dimensionality can be reduced to $(n_s+1) \times
(n_s+1$). An additional reduction of the dimensionality of the problem
can be achieved by introducing certain approximations. We will follow
here earlier studies \cite{vakarin1997, kalyuzhnyi1998, duda1998,
kalyuzhnyi2003, kalyuzhnyi2011}, that were carried out for different
versions of models with site-site bonding and utilize the analogue
of the so-called ``ideal network approximation'' (INA) combined with the
orientationally averaged version of the multi-density OZ equation.
According to these approximations it is assumed that

\begin{equation}
c_\alpha (1) = 0 ~~~ {\rm and} ~~~ \varepsilon_\alpha (1) = 0, ~~~~ 
{\rm for} ~~~ |\alpha|>1
\label{idealc}
\end{equation}
where $| \alpha |$ denotes the cardinality of set
$\alpha$. Further the OZ equation (\ref{OZoper}) and the closure
relations are expressed in terms of the orientationally averaged
partial correlation functions $h_{\alpha\beta}(r)$ and
$c_{\alpha\beta}(r)$. We will henceforward drop the ubiquitous
explicit argument `1', since we consider a uniform system.
Within the INA and taking into
account the equivalence of the $n_s$ sites, the dimensionality of the
OZ equation can finally be reduced to $2\times 2$.

As a consequence of the INA (\ref{idealc}),
all correlation functions which involve particles with more than one
bonded site are neglected. However, this does not mean that
correlations between particles in all possible bonded states are
neglected.  Instead, they are accounted for via the convolution terms
in the right-hand side of the OZ equation due to the appearance of the
density parameters $\hat{\sigma}$ introduced in Equation
(\ref{eq:sigma}). In a certain sense this approximation is similar
(but not equivalent) to the approximation utilized in the
thermodynamic perturbation theory of Wertheim \cite{wertheim1986b,
wertheim1987td}. 

In the following, boldfaced symbols collect partial correlation functions.
Replacing the angular dependent correlation functions ${\bm h}(1, 2)$
and ${\bm c}(1,2)$ in the OZ equation (\ref{OZoper}) by their
orientationally averaged counterparts, ${\bm h}(r)$ and ${\bm c}(r)$,
we arrive at

\begin{equation}
{\bm h}(r_{12}) = {\bm c}(r_{12})+ 
\int\;{\bm c}(r_{13}){\bm\sigma}{\bm h}(r_{32})\;d {\bf r}_3,
\label{eq:OZIPC}
\end{equation}
where

\begin{equation}
{\bm h}(r)=
\left( \begin{array}{cc}
h_{00}(r) & h_{01}(r) \\
h_{10}(r) & h_{11}(r)
\end{array}
\right) ~~~
{\bm c}(r)=
\left( \begin{array}{cc}  
c_{00}(r) & c_{01}(r) \\
c_{10}(r) & c_{11}(r)
\end{array} \right) ~~~
{\bm\sigma}=
\left( \begin{array}{cc}
\rho & n_s\sigma_{n_s-1} \\
n_s\sigma_{n_s-1} & n_s(n_s-1)\sigma_{n_s-2}
\end{array} \right).
\label{eq:h_c_sigma}
\end{equation}
In equation (\ref{eq:h_c_sigma}) the following notation is used
\[
h_{0K}(r)\equiv h_{01}(r),\;\;\;\;\;
h_{KL}(r)\equiv h_{11}(r),
\]
\[
\sigma_{\Gamma - K}\equiv \sigma_{n_s-1},\;\;\;\;\;
\sigma_{\Gamma - K - L}\equiv \sigma_{n_s-2}.
\]

The AHNC closure (\ref{gfinal}) with $\hat{E}(1, 2)=0$ takes the form

\begin{eqnarray} \nonumber
c_{00}(r) & = & g_{00}(r)-t_{00}(r)-1, \\ \nonumber
c_{01}(r) & = & g_{00}(r)\left[t_{01}(r)+f_{01}(r)\right]-t_{01}(r), \\ \nonumber
c_{10}(r) & = & g_{00}(r)\left[t_{10}(r)+f_{10}(r)\right]-t_{10}(r), \\ 
c_{11}(r) & = & g_{00}(r)\left[t_{11}(r)+t_{01}(r)t_{10}(r) 
+f_{01}(r)t_{10}(r)+f_{10}(r)t_{01}(r) \right]-t_{11}(r) ,
\label{c_hnc_11}
\end{eqnarray}
while the APY closure (\ref{PY}) reads as

\begin{eqnarray} \nonumber
c_{00}(r)& = & f_{\rm ref}^{(0)}(r)\left\{t_{00}(r)+1\right\}, \\ \nonumber
c_{01}(r)& = & e_{\rm ref}^{(0)}(r)\left\{t_{01}(r)+\left[t_{00}(r)+1\right]
f_{01}(r)\right\}-t_{01}(r), \\ \nonumber
c_{10}(r) & = & e_{\rm ref}^{(0)}(r)\left\{t_{10}(r)+\left[t_{00}(r)+1\right]
f_{10}(r)\right\}-t_{10}(r), \\ 
c_{11}(r) & = & e_{\rm ref}^{(0)}(r)\left\{t_{11}(r)+
t_{01}(r)f_{10}(r)+t_{10}(r)f_{01}(r) \right\}-t_{11}(r),
\label{c_py}
\end{eqnarray}
where 

$$
e_{\rm ref}^{(0)}(r) = \exp{\left[-\beta U_{00}(r)\right]} ~~~~~~
f_{\rm ref}^{(0)}(r)=e_{\rm ref}^{(0)}(r)-1 ~~~~~~
g_{00}(r)=e_{\rm ref}^{(0)}(r)\exp{\left[t_{00}(r)\right]}
$$
and
$$
t_{ij}(r)=h_{ij}(r)-c_{ij}(r) .
$$

Note that in the expressions (\ref{c_hnc_11}) and (\ref{c_py})
the term $F_{11}(1, 2)=e_{\rm ref}(1, 2)f_{01}(1,2)f_{10}(1, 2)$,
which takes into account contributions due to double bonding, 
has been dropped and that the Boltzmann factor for the
reference potential, $e_{\rm ref}(1, 2)$, has been approximated by the
Boltzmann factor $e_{\rm ref}^{(0)}(r)$ for the potential $U_{00}(r)$,
which is acting between the centers of the colloidal particles.

Finally, in order to obtain a closed set of equations the relations
between the density parameters $\sigma_i$, introduced in Equation
(\ref{eq:h_c_sigma}), and the pair distribution functions $g_{ij}(r)$
have to be specified. This can be achieved by combining equations
(\ref{eq:rhoexp}), (\ref{eq:sigmainv}), (\ref{cusual}) and
(\ref{idealc}). One obtains the following relations

\begin{equation}
n_s\rho X^2\int e_{\rm ref}^{(0)}(r)f_{10}(r)y_{01}(r) dr +
X \left[ \rho \int e_{\rm ref}^{(0)}(r)f_{10}(r)y_{00}(r) dr + 1 \right]-1 = 0,
\end{equation}

\begin{equation}
\sigma_{n_s-2}=\rho X^2 ~~~~ {\rm with} ~~~ X = X_{\Gamma - K}
= \sigma_{n_s-1}/\rho.
\label{densities}
\end{equation}
where $X$ is the fraction of particles where the patch (site) $K$
is not bonded.

For the partial cavity correlation functions, $y_{ij}(r)$, defined in
Equations (\ref{eq:yNE}) and (\ref{eq:t}), we find within the AHNC
approximation

\begin{eqnarray} \nonumber
y_{00}(r) & = & \exp{\left[t_{00}(r)\right]} \\ \nonumber
y_{01}(r) & = & y_{00}(r)t_{01}(r) \\ \nonumber
y_{10}(r) & = & y_{00}(r)t_{10}(r) \\ 
y_{11}(r) & = & y_{00}(r) \left[ t_{01}(r)t_{10}(r) + t_{11}(r) \right]
\label{yAHNC}
\end{eqnarray}
and within the APY approximation

\begin{equation}
y_{ij}(r)=t_{ij}(r)+\delta_{i0}\delta_{j0} .
\label{yPY}
\end{equation}

The use of the orientationally averaged version of the OZ relation, specified in Equation~(\ref{eq:OZIPC}), introduces an additional approximation. The orientational averaging, originally proposed in Refs.~\cite{vakarin1997, kalyuzhnyi1998, duda1998, kalyuzhnyi2003, kalyuzhnyi2011} might seem to be a crude approximation since the correlation functions entering the OZ equation~(\ref{eq:OZIPC}) and the closure relations~(\ref{c_hnc_11}) and~(\ref{c_py}) do not display an explicit dependence on the orientational degrees of freedom. Nevertheless we note that, even though APY does not account directly for the repulsion between patches, it correctly takes into account the major effect due to the patch-patch repulsion, i.e. restricting the appearance of the double bonds between the particles. According to previous studies \cite{wertheim1987, kalyuzhnyi1991, duda1998, kalyuzhnyi2011} the orientationally averaged theory is able to provide accurate results for both the structural and the thermodynamic data of related systems.

Together with either the AHNC (\ref{c_hnc_11}) or the APY (\ref{c_py}) closure relations, Equations (\ref{eq:OZIPC}) and (\ref{densities}) form a closed set of equations that has to be solved.

Finally, the total pair distribution function $g(r)$ is obtained from
the partial distribution functions $g_{ij}(r)$ via the following relation

\begin{equation}
g(r)=g_{00}(r)+n_sXg_{01}(r)+n_sXg_{10}(r)+n_s^2X^2g_{11}(r).
\label{g_tot}
\end{equation}

Now we derive the expressions needed to calculate thermodynamic properties; they are based on the solution of the OZ equation (\ref{eq:OZIPC}). In our derivation of the expressions for the internal energy $E$ and for the (virial) pressure  $p^{\rm v}$ of
the IPC model in terms of the partial distribution functions
$g_{ij}(r)$ we used a scheme that was developed for a model with one
attractive site \cite{wertheim1986dim, wertheim1987, kalyuzhnyi1991}.
We start from the following general expressions

\begin{equation}
{E\over V} = {1\over 2} \int \rho(1, 2) U(1, 2) 
d {\bf r}_{12} d\Omega_1  d\Omega_2 =
-{1\over 2} \int {\rho(1,2) \over e(1,2)}
{\partial e(1, 2) \over \partial \beta} 
d {\bf r}_{12} d\Omega_1  d\Omega_2
\label{E}
\end{equation}
and 

\begin{equation}
{\beta p^{\rm v} \over \rho} = 1 - {\beta\over 6\rho} \int \rho(1, 2)
{\bf r}_{12} {\nabla}_2 U(1, 2) d{\bf r}_{12} d\Omega_1 d\Omega_2 = 1
+ {1\over 6\rho} \int {\rho(1, 2)\over e(1, 2)} {\bf r}_{12}
{\nabla}_2 e(1, 2) d{\bf r}_{12} d\Omega_1 d\Omega_2 ,
\label{P}
\end{equation}
where ${\bf r}_{12} = {\bf r}_1 - {\bf r}_2$ and $\rho(1, 2) = \rho(1)
g(1, 2) \rho(2)$ is the pair density.

The derivatives in the above relations can be rewritten as

\begin{equation}
{\partial e(1, 2) \over \partial \beta} =
{\partial e_{\rm ref}(1,2) \over \partial \beta} +
\sum_K \left[ {\partial F_{K, 0}(1, 2) \over \partial \beta}
+ {\partial F_{0K}(1, 2) \over \partial \beta} \right]
\label{dbet}
\end{equation}
and

\begin{equation}
{\nabla}_2e(1, 2)=
{\nabla}_2e_{\rm ref}(1, 2)+\sum_K\left({\nabla}_2 F_{K0}(1, 2)
+{\nabla}_2 F_{0K}(1, 2)\right) .
\label{dr}
\end{equation}

We now substitute these expressions into Equations (\ref{E}) and
(\ref{P}) and perform a resummation of the diagrams representing
$\rho(1, 2)/e(1, 2)$ in terms of the activity $z$. Within the INA, and
with a subsequent replacement of the orientationally dependent quantities
by their averaged counterparts, we find the following expression

\begin{eqnarray}
{\beta E\over V} & = & 2\pi\beta\int g(r)U_{00}(r)r^2dr \\ \nonumber
& & - 2\pi\beta n_sX\int e_{\rm ref}^{(0)}(r)\left\{
\Big[y_{00}(r)+n_sXy_{01}(r)\Big]{\partial f_{10}(r)\over\partial\beta} 
+\Big[y_{00}(r)+n_sXy_{10}(r)\Big]
{\partial f_{01}(r)\over\partial\beta}\right\}r^2dr
\label{Efin}
\end{eqnarray}
and

\begin{eqnarray}
{\beta p^{\rm v} \over\rho} & = & 
1-{2\pi\over 3}\beta\rho\int g(r){\partial U_{00}(r)\over\partial r} \\
\nonumber
& & +{2\pi\over 3}\rho n_sX\int e_{\rm ref}^{(0)}(r)\left\{
\Big[y_{00}(r)+n_sXy_{01}(r)\Big]{\partial f_{10}(r)\over\partial r}
+\Big[y_{00}(r)+n_sXy_{10}(r)\Big]{\partial f_{01}(r)\over\partial r}\right\}
r^3dr .
\label{Pv}
\end{eqnarray}
These expressions are valid for $n_s$ equivalent patches and can be
used in combination with any approximate closure relation. 

In addition we also present the expression for the pressure calculated
via the compressibility route, $p^{\rm c}$; it can be obtained using
the APY closure relation (\ref{c_py}) and following a
scheme developed by Wertheim \cite{wertheim1986b, wertheim1987}

\begin{eqnarray}
{\beta p^{\rm c}\over\rho} & = & 
1-{2\pi\over\rho}\int_0^\infty
\left[{\tilde \bsig}{\bf C}(r){\tilde \bsig}\right]_{00}r^2dr \\
& & +{1\over 2\pi^2\rho}\sum_{\ell=0}^{n_s}\int_0^\infty
\left\{{{\hat\lambda}_\ell^2(k)
\over 2(1-{\hat\lambda}_\ell(k))}+
{\hat\lambda}_\ell(k)+\ln{\left[1-{\hat\lambda}_\ell(k)\right]}
\right\}k^2dk.
\label{Pc}
\end{eqnarray}
Here ${\tilde \bsig}$ and ${\bf C}(r)$ are matrices with the following
elements ($i,j=1,\ldots,n_s$)

\begin{equation}
\left[{\tilde \bsig}\right]_{00}=\rho,\;\;\;
\left[{\tilde \bsig}\right]_{0i}=\left[{\tilde \bsig}\right]_{i0}=\sigma_{n_s-1},\;\;\;
\left[{\tilde \bsig}\right]_{ij}=(1-\delta_{ij})\sigma_{n_s-2},
\end{equation}

\begin{equation}
\left[{\bf C}(r)\right]_{00}=c_{00}(r),\;\;\;
\left[{\bf C}(r)\right]_{0i}=c_{01}(r),\;\;\;
\left[{\bf C}(r)\right]_{i0}=c_{10}(r),\;\;\;
\left[{\bf C}(r)\right]_{ij}=c_{11}(r) ;
\end{equation}
the ${\hat\lambda}_{i}(k)$ denote the eigenvalues of the matrix
${\hat{\bf C}}(k){\tilde \bsig}$, with the elements of the matrix
${\hat{\bf C}}(k)$ being the Fourier transforms of the corresponding
elements of the matrix ${\bf C}(r)$.

Note that in the above expressions the reference potential $U_{\rm
  ref}(1, 2)$ has been approximated by the potential $U_{00}(r)$
  acting between the colloidal centers.

\subsection{The reference hypernetted-chain approach}
\label{subsec:rhnc}

This approach is based on the OZ equation for
orientationally dependent correlation functions~\cite{HeM},

\begin{equation}
h(1,2) = c(1,2) + \frac{\rho}{4\pi} \int c(1,3) h(3,2) d3 ,
\label{eq:OZ}
\end{equation}
where $h(1,2) = g(1,2) - 1$ is the total correlation function and $c(1, 2)$ the direct correlation function. 

This relation is complemented by the RHNC closure
\begin{equation}
g(1,2) = \exp \left [ -\beta U(1,2) + N(1,2) + E(1,2) \right ],
\label{eq:hnc}
\end{equation}
where $U(1, 2)$ is the potential, $E(1, 2)$ is the bridge function, and $N(1,2) = h(1,2) - c(1,2)$ is the indirect correlation function.
The above relations are formally exact; however, since $E(1, 2)$ is known only in terms of a complex diagrammatic representation one has to resort to approximate schemes introduced below.

A route to solve equations (\ref{eq:OZ}) and (\ref{eq:hnc}) was originally proposed by Lado~\cite{Lado}; ever
since this method was steadily extended and improved.
Specific details of the last version are summarized in Ref.~\cite{Giacometti}.
For this contribution we have adapted the most recent version of the code
to our system of IPCs. In the following we briefly sketch the algorithm.

In Lado's approach Equation~(\ref{eq:OZ}) is solved by projecting the correlation functions onto
spherical harmonics, e.g., for a generic function $f(1,2)$,
\begin{equation}
f (1, 2) = f(r,\theta_1,\theta_2,\varphi) = 4 \pi \sum_{\ell_1=0}^\infty
\sum_{\ell_2=0}^\infty \sum_{m=-M}^M f_{\ell_1 \ell_2 m}(r)
Y_{\ell_1}^m(\theta_1,\varphi_1) Y_{\ell_1}^{-m}(\theta_2,\varphi_2), 
\label{eq:expansion}
\end{equation}
where $M = \min(\ell_1,\ell_2)$, $\theta_i$ is the angle between
${\bf r}_{12}$, the vector joining the two particle centers,
and $\boldsymbol{\omega}_i$, the orientational unit vector of particle $i$,
and $\varphi=\varphi_2-\varphi_1$ is the angle between $\boldsymbol{\omega}_1$ and
$\boldsymbol{\omega}_2$ in the plane orthogonal to ${\bf r}_{12}$.
Via a Fourier transform with respect to $r$, we can transform Equation~(\ref{eq:OZ}) into a
linear system of equations involving the respective coefficients.
In contrast, all functions involved in the non-linear closure relation~(\ref{eq:hnc}) have to be treated as full functions of their spatial and orientational variables.

The numerical program follows an iterative procedure that can be divided into two steps.
In the first one, we start from an initial guess for $N(1,2)$, and so
the closure relation (\ref{eq:hnc}) is solved for $c(1,2)$.
The bridge function we used is $E_{\rm HS}(r; D^*)$, i.e. the bridge of an HS fluid calculated via the parametrization proposed in~\cite{VerletWeiss}; the parameter $D^*$ is an effective HS diameter that differs from the diameter of the IPC particles, and for which at the beginning of the iterative procedure a suitable guess is assumed.
In the second step, $c(1,2)$ is expanded in terms of the $c_{\ell_1 \ell_2 m}(r)$, which are given by a transformation that is inverse to the one specified in Equation~(\ref{eq:expansion}), namely
\begin{equation}
 f_{\ell_1 \ell_2 m}(r)= \frac{1}{(4\pi)^2} \int  [Y_{\ell_1}^{m}(\theta_1,\varphi_1) Y_{\ell_2}^{-m}(\theta_2,\varphi_2)]^*
 f(r,\theta_1,\theta_2,\varphi)  \sin\theta_1 \di\theta_1 \sin \theta_2 \di\theta_2 \di \varphi_1 \di \varphi_2;
\end{equation}
here the star denotes complex conjugation and the angular integrals are carried out using Gaussian quadratures.
These coefficients are are then Fourier transformed so that Equation~(\ref{eq:OZ}) can be solved for
$\tilde{N}_{\ell_1 \ell_2 m}(k) = \tilde{h}_{\ell_1 \ell_2 m}(k) - \tilde{c}_{\ell_1 \ell_2 m}(k)$.
Finally, the new $N_{\ell_1 \ell_2 m}(r)$ and the old $g_{\ell_1 \ell_2 m}(r)$
are used to calculate the new $c(1,2)$ via its coefficient functions
\begin{equation}
  c_{\ell_1 \ell_2 m}(r) = g_{\ell_1 \ell_2 m}(r) - \delta_{\ell_1 0}\delta_{\ell_2 0}\delta_{m 0} - \gamma_{\ell_1 \ell_2 m}(r).
\end{equation}
These two steps constitute an iteration loop.

When convergence is achieved (i.e., the difference in the correlation
functions of two subsequent steps differ less than a small threshold
value, in our case typically $10^{-5}$) for a particular value of
$D^*$, then this parameter is modified and the iterative scheme is
repeated, until the free energy of the system, which is convex with
respect to $D^*$, has been minimized.

The initial guess for $g(1,2)$ and $N(1,2)$ stems, whenever possible, from a previous solution of 
the problem for a neighboring state point; otherwise the program can resort to the parametrization of the isotropic HS counterparts of $g(1,2)$ and $N(1,2)$, proposed in Ref. \cite{VerletWeiss}.

Once the algorithm has converged, the internal energy and the pressure are obtained from the following equations~\cite{Giacometti}
\begin{equation}
  \frac{E}{N} = \frac{\rho}{2}\int d{\bf r}_{12} \langle g(1,2)U(1,2)\rangle_{\boldsymbol{\omega}_1 \boldsymbol{\omega}_2 }
\end{equation}
and
\begin{equation}
  p = \rho k_B T - \frac{\rho^2}{6}\int d{\bf r}_{12} \left \langle g(1,2) r_{12} \frac{\partial}{\partial r_{12}} U(1,2) \right \rangle_{\boldsymbol{\omega}_1 \boldsymbol{\omega}_2 }
\end{equation}
where $\langle \dots \rangle_{\boldsymbol{\omega}_i} =
\frac{1}{4\pi}\int \dots \di \boldsymbol{\omega}_i$  denotes an angular average over the orientation of the two particles.

\subsection{The Barker-Henderson thermodynamic perturbation theory}
\label{subsec:BH-TPT}

In recent work~\cite{Gogelain} the original Barker-Henderson
thermodynamic perturbation theory (BH-TPT)~\cite{bhoriginal}
was extended to systems with anisotropic potentials.
Within this framework the Helmholtz free energy, $A$, is
calculated as a truncated series expansion, starting from the free
energy of the reference system, which in our case is the HS one, namely
\begin{equation}
\frac{\beta A}{N} = \frac{\beta A_{\rm HS}}{N} + \sum_i \frac{\beta A_i}{N} .
\label{eq:BH-TPT}
\end{equation}
For the hard-sphere contribution, $A_{\rm HS}$, we use the
Carnahan-Starling equation of state \cite{CarnahanStarling}. The first
two terms of the above series expansion are given by \cite{Gogelain}
\begin{equation}
\frac{\beta A_1}{N} = 2\pi\rho \int_{D}^{D+\delta} g_{\rm HS}(r) 
\langle \beta U(r, \boldsymbol{\omega}_1 \boldsymbol{\omega}_2)
\rangle_{\boldsymbol{\omega}_1 \boldsymbol{\omega}_2}
dr 
\end{equation}
and
\begin{equation} 
\frac{\beta A_2}{N} = - \left( \frac{\pi \rho}{\tilde{\chi}} \right)
\int_{D}^{D+\delta} g_{\rm HS}(r) \langle [\beta
  U(r, \boldsymbol{\omega}_1 \boldsymbol{\omega}_2]^2
\rangle_{\boldsymbol{\omega}_1 \boldsymbol{\omega}_2}
dr .
\end{equation}
For the quantity $\tilde \chi$, defined as
\begin{equation} 
\tilde{\chi} = \frac{6}{\pi}\dede{}{\rho} 
\left( \frac{\beta p}{\rho} \right)_{\rm HS}
\end{equation}
we use again the Carnahan-Starling equation of state, while
for the HS radial distribution function, $g_{\rm HS}(r)$, we
rely on the Verlet-Weis parametrization \cite{VerletWeiss}.

Having computed the free energy up to second order via
Equation (\ref{eq:BH-TPT}), we obtain the pressure according to
\begin{equation} 
\frac{\beta p}{\rho} = \rho \dede{}{\rho} 
\left( \frac{\beta A}{N} \right).  
\end{equation}
The BH-TPT does not provide any expressions to directly calculate the internal energy.

\subsection{Monte Carlo simulations}
\label{subsec:mc}

We perform Monte Carlo (MC) simulations both in the $NVT$ as well as in the $NpT$ ensemble~\cite{frenkelsmit}. 
In the canonical ensemble, each MC step consists of $N$ trial particle moves, where the acceptance rule is imposed via the Metropolis algorithm. A particle move is defined as both a particle displacement in each of the Cartesian directions by a random quantity, uniformly distributed within $[- \delta r, + \delta r]$, and a rotation of the particle around a random spatial axis by a random angle, uniformly distributed within $[- \delta \vartheta, + \delta \vartheta]$ with $\delta \vartheta = 2 \delta r/ D$. 
In the isobaric-isothermal ensemble, each MC move consists of $N$ trial particle moves as defined above and one trial change in the volume; the latter one is an attempt to change the volume of the cubic box by a random quantity, uniformly distributed within $[- \delta V, + \delta V]$, see Ref.~\cite{frenkelsmit}. During the equilibration runs, the values of $\delta r$ and $\delta V$ are allowed to change in order to guarantee an acceptance ratio of the corresponding moves between $30 \%$ and $50 \%$. Successive sampling runs with constant values of $\delta r$ and $\delta V$ extend over $10^6$ MC steps in the $NVT$ ensemble and $3 \cdot 10^6$ MC steps in the $NpT$ ensemble.

In the $NVT$ ensemble, we evaluate the {\it structural} properties of the system at hand. We consider ensembles of $N = 1000$ particles in a cubic volume with edge lengths $L \approx 17 D$ and $L \approx 13 D$; the corresponding number densities are $\rho D^3 = 0.20$, and $0.45$, respectively. For each system, we choose four temperatures, namely $T^*=0.5,\;0.32,\;0.23,\;0.18$ (in units of $k_{\rm B}T/\epsilon_{\rm m}$, where $\epsilon_{\rm m}$ is the minimum of the pair interaction energy, see Table~\ref{tab:model-parameters}). For each state point, we determine the radial distribution function, $g(r)$, the average number of bonds per particle, $Q$, and the average number of particles that form $n_b$ bonds with other particles in the system, $N(n_b)$. All these quantities are averaged over 2000 independent configurations.

In the $NpT$ ensemble, we consider systems of $N=1000$ particles at the same four temperatures specified above and we consider eleven pressure values, ranging from $p^*=0.01$ to $0.50$ (in units of $pD^3/\epsilon_{\rm m}$). At each state point, we determine the thermodynamic properties of these systems, namely the internal energy per particle $E/N$ and the equilibrium number density, $\rho$. Both quantities are averaged over 2000 independent configurations.

\section{Results and Discussion}
\label{sec:results_discussion}

\begin{figure}[h] 
\centering
\includegraphics[width=\columnwidth, clip=true]{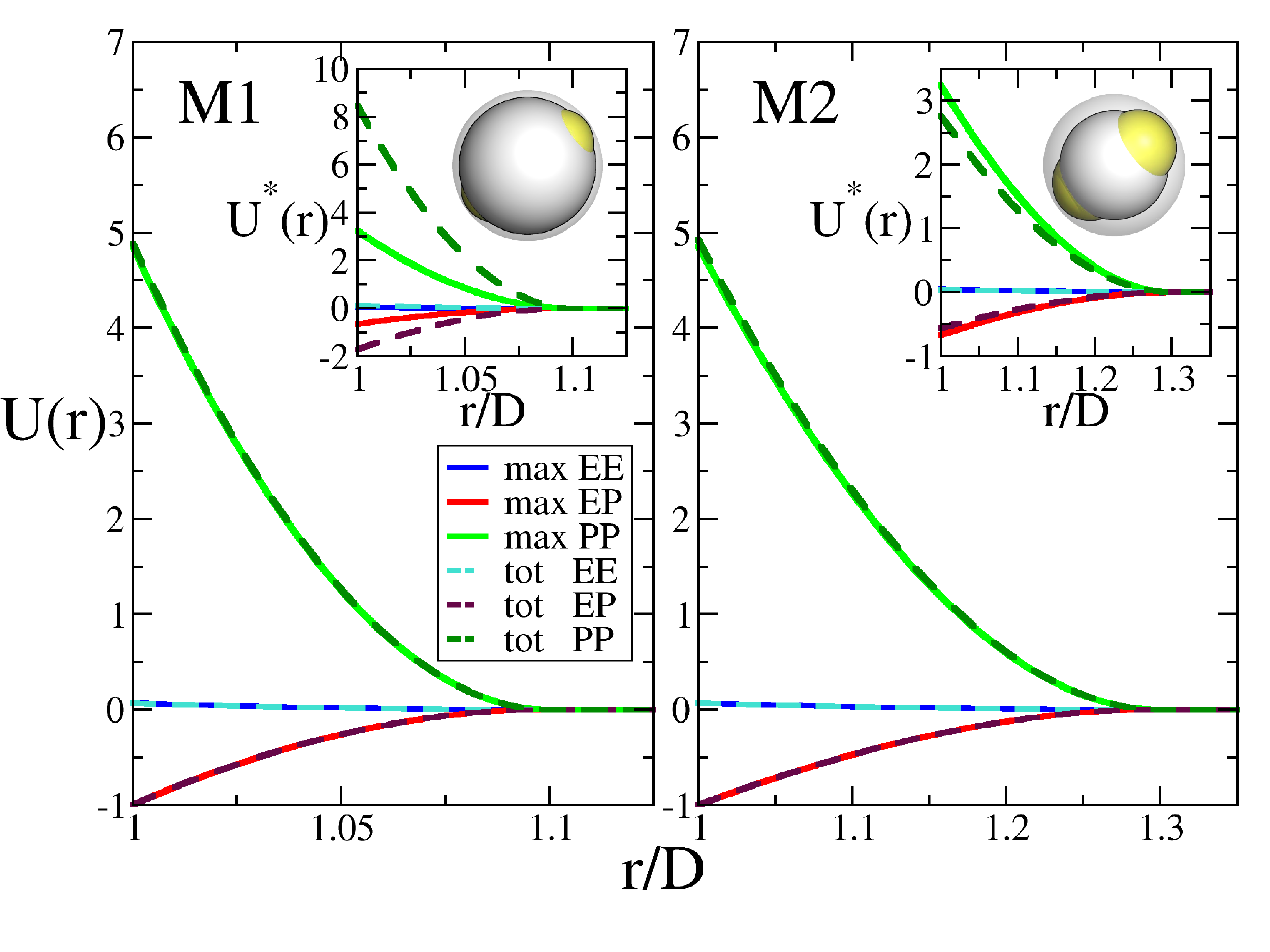}
\caption{(Color online) The main panels show the normalized interaction 
  energy $U(r)$ in units of $k_BT$ between two IPCs as a function
  of their distance $r$; left panel -- model M1, right
  panel -- model M2. Three particular particle configurations have
  been considered, referred to as polar-polar (PP; dark and light
  green), equatorial-equatorial (EE; dark and light blue), and
  equatorial-polar (EP; dark and right red).
  The insets display the non-normalized pair energy $U^*(r)=U(r)\epsilon_{\rm m}$, which correspond to the pair potentials shown in panels (a) and (c) of Figure~7 in Ref.~\cite{bianchi2011}.
  In all the graphs, continuous and dashed lines correspond to the two different coarse-graining procedures put
  forward in Ref.~\cite{bianchi2011} and termed there ``tot''- and ``max''-schemes, respectively.
  It is worth noting that, once normalized, potentials obtained via both the ``tot'' and the ``max'' routes coincide.
  Schematic representations of the two models are shown, where particle size and patch extent are drawn to scale.}
\label{fig:potentials}
\end{figure}

We have studied the properties of two selected systems of IPCs, denoted as M1 and M2 and specified in Table~\ref{tab:model-parameters} via their model parameters. The corresponding pair interactions are shown in Figure \ref{fig:potentials} for three characteristic particle configurations.
The difference between the two models lies in both the interaction range and the patch size. In model M1 we have $\delta =$ 20\% of $D/2$ and $\gamma \simeq 22^\circ$, while in model M2 we have $\delta =$ 60\% of $D/2$ and $\gamma \simeq 43^\circ$.

In the following we show and discuss structural and thermodynamic data
obtained for the two model systems in MC simulations (obtained in the
$NVT$ or $NpT$ ensembles) and via the APY approach; where applicable (i.e., if
convergence could be achieved) results from RHNC and BH-TPT are added.
The AHNC results are not shown in the following since they were found to be very similar to the APY ones; further the APY closure was found to converge in a broader region of the phase diagram.

\subsubsection{Structural properties}

We start our discussion with a comparison of the results obtained for
the pair distribution function, $g(r)$; MC data were obtained in $NVT$
simulation runs. A thorough check of $g(r)$ is of particular relevance
as this function forms the basis of the subsequent calculation of thermodynamic quantities, such as energy
and pressure. In addition, we have analysed and compared results for the average number of bonds per particle, $Q$, as obtained from the simulation and from the APY approach; while in simulation this quantity is obtained by simple counting of bonds, in the theoretical approach it is calculated via the following expression:

\begin{equation}
Q = \pi \rho \int_D^{D+\delta} 
\left[ g(r) - e_{\rm ref}^{(0)}(r)y(r) \right]r^2dr,
\label{alpha}
\end{equation}
where $y(r)=y_{00}(r)+n_sXy_{01}(r)+n_sXy_{10}(r)+n_s^2X^2y_{11}(r)$.

\begin{figure}[h] 
\centering
\includegraphics[width=\columnwidth, clip=true]{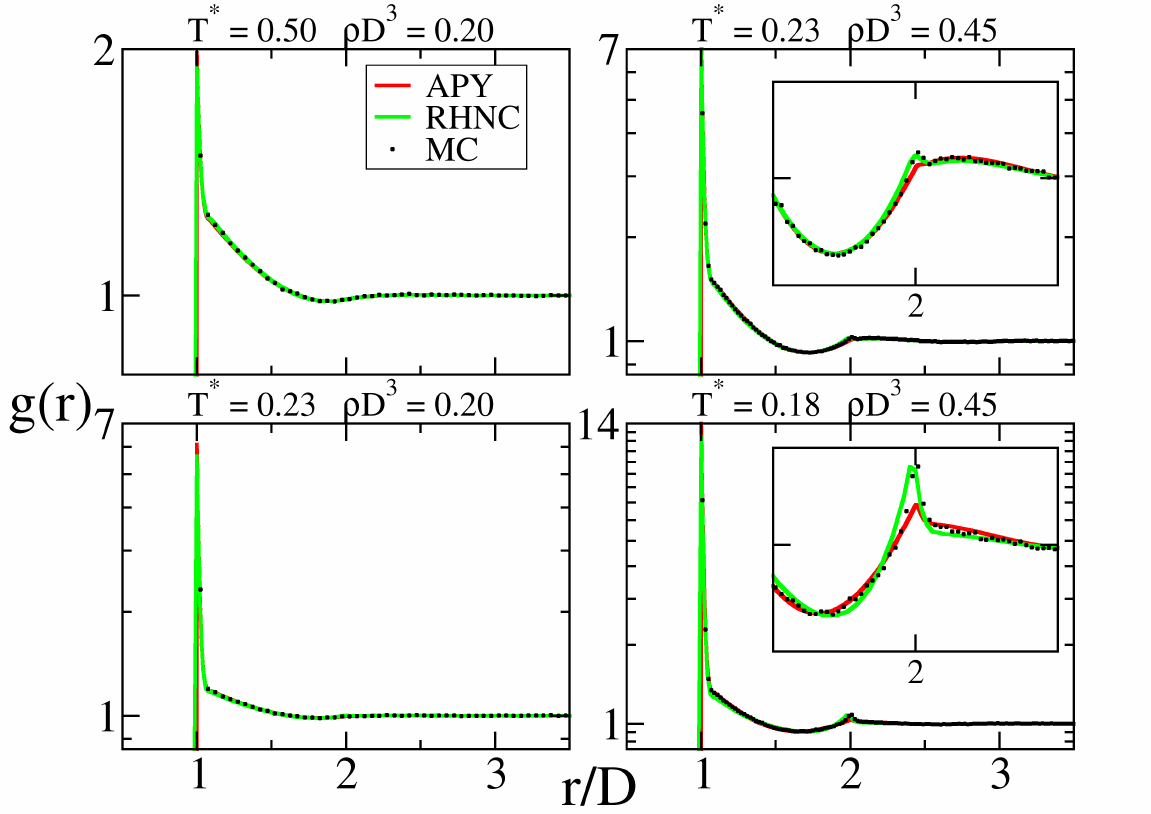}
\caption{(Color online) Pair correlation functions $g(r)$ for model M1 at
  selected state points (as labeled); note the logarithmic scale of
  the vertical axis. Symbols correspond to MC simulation data, while
  lines represent either APY (red) or RHNC (green) results. Insets
  show enlarged views around the second peak of the $g(r)$ when
  appropriate.}
\label{fig:gr-m1}
\end{figure}

In Figure \ref{fig:gr-m1} we show results for $g(r)$ of system
M1. In the left column we display data of state points
characterized by a low density ($\rho D^3 = 0.2$), while the panels in
the right column show data at high density ($\rho D^3 = 0.45$);
temperature decreases as we proceed from the top to the bottom
panels.

In the low-density, high-temperature state point (upper-left panel),
$g(r)$ shows a main peak of relative moderate height and a very flat,
hardly visible maximum at the second nearest neighbour distance
($r \sim 2D$); simulations and all theoretical data
are in excellent agreement. As we decrease the temperature from $T^* =
0.50$ to $T^* = 0.23$, the contact value of $g(r)$ increases
substantially; the agreement between the different data sets is still very good, only
small discrepancies are observed at contact (i.e., at $r = D$).

As we proceed to the higher density, we observe a substantial change in
the shape of $g(r)$: while we still observe the pronounced peak at
contact, a distinctive, characteristic maximum at the second nearest
neighbour distance emerges (see insets of the panels). For both
high density state points these features are reproduced on a
qualitative level by all approaches; on a more
quantitative level we observe that differences between APY data and
simulation results are only found in the immediate vicinity of the
second peak in $g(r)$.

\begin{figure}[h] 
\centering
\includegraphics[width=\columnwidth, clip=true]{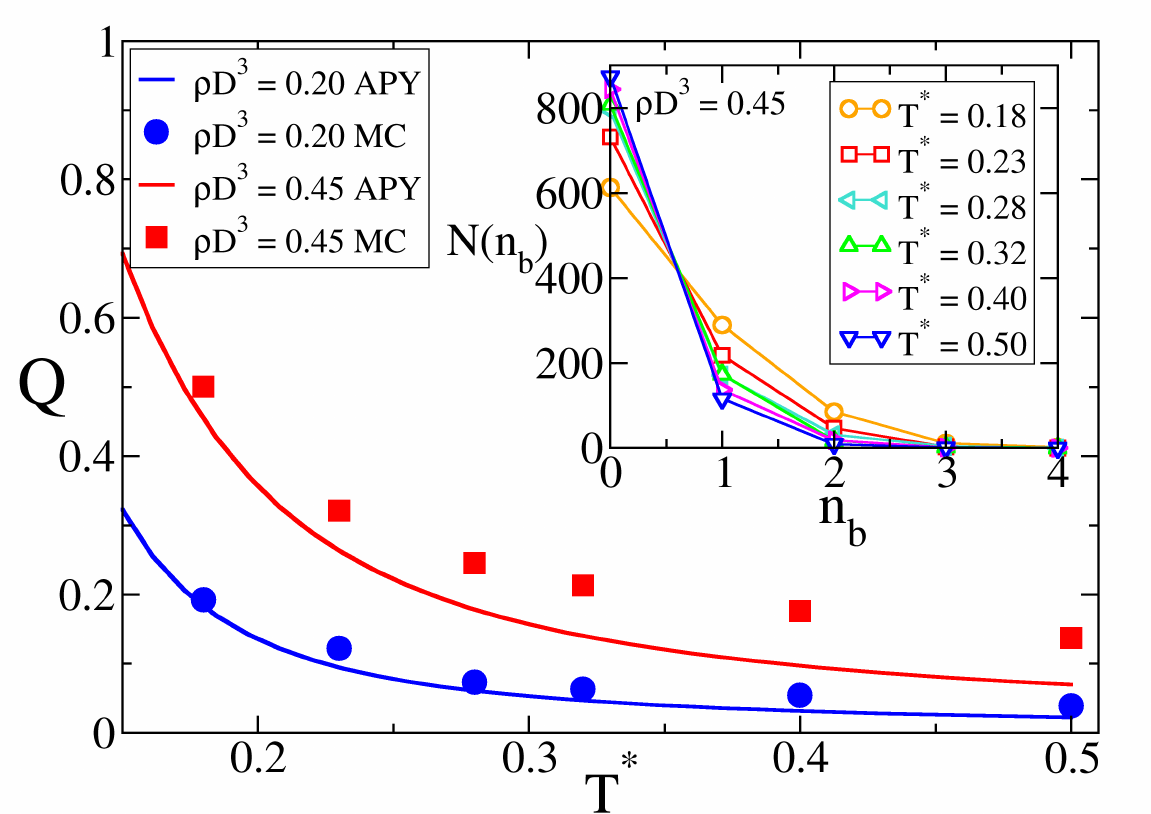}
\caption{(Color online) Average number of bonds per particle $Q$ versus temperature 
  for system M1 at two different densities (as labeled); for
  the definition of $Q$ see text. Symbols correspond to MC data while
  lines represent APY results. The inset displays simulation data for the average number of 
  particles with $n_{\rm b}$ bonds, $N(n_{\rm b})$, versus the number of bonds 
  for selected temperatures at density $\rho D^3 = 0.45$ (as labeled).}
\label{fig:X-m1}
\end{figure}

In Figure \ref{fig:X-m1} we display the average number of bonds per
particle, $Q$, as a function of temperature for model
M1. Along the low-density branch the agreement between
the different sets of data is very satisfactory; in contrast, at $\rho D^3
= 0.45$ discrepancies become substantial at intermediate and higher
temperatures. 
In the inset of this figure we
display simulation results for the average number of particles that form $n_b$
bonds, $N(n_b)$, for both high density states at different
temperatures. These data confirm the expected trend:
while at higher temperature particles are preferentially isolated,
at lower temperature most of them is in bonded states.

\begin{figure}[h] 
\centering
\includegraphics[width=\columnwidth, clip=true]{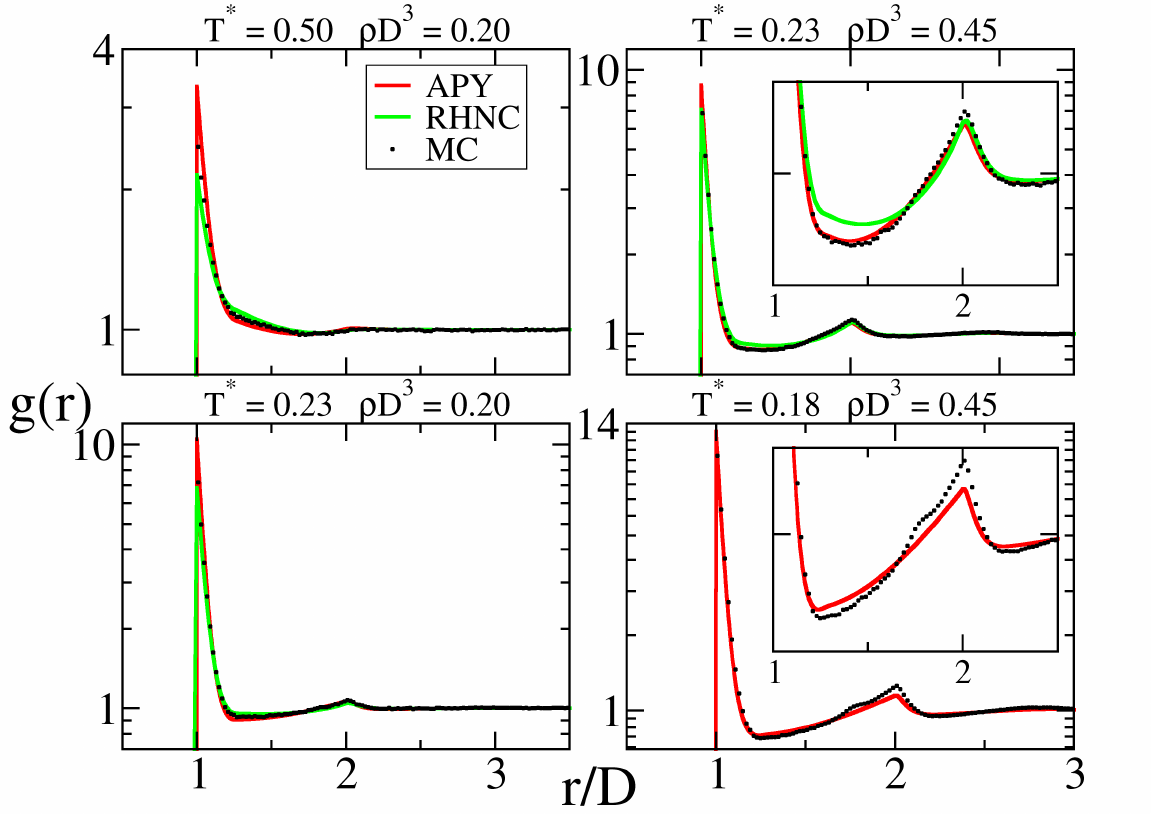}
\caption{(Color online) Pair correlation functions $g(r)$ for model M2 at
  selected state points (as labeled); note the logarithmic scale of
  the vertical axis. Symbols correspond to MC simulation data, while
  lines represent either APY (red) or RHNC (green) results. Insets
  show enlarged views around the second peak of the $g(r)$ when
  appropriate.}
\label{fig:gr-m2}
\end{figure}

In Figure \ref{fig:gr-m2} we show results for $g(r)$ of model
M2. Again, density increases from left to right while
temperature decreases from top to bottom. The fact that the
interaction range is now substantially broader than in
model M1 represents a severe and thus very stringent
test for the reliability of the theoretical approaches: except for the
high-density, low-temperature state, both APY and RHNC provide
converged results. In contrast to model M1,
discrepancies between the different approaches (simulation, APY and
RHNC) are now visible; still, the characteristic features of $g(r)$,
i.e., the pronounced peak at contact and the emergence of a maximum at
the second nearest neighbour distance are reproduced throughout at a
qualitative level. We observe that in particular the contact values
and the features of the second maximum are very sensitive to the
respective approaches (see in particular the insets).

\begin{figure}[h] 
\centering
\includegraphics[width=\columnwidth, clip=true]{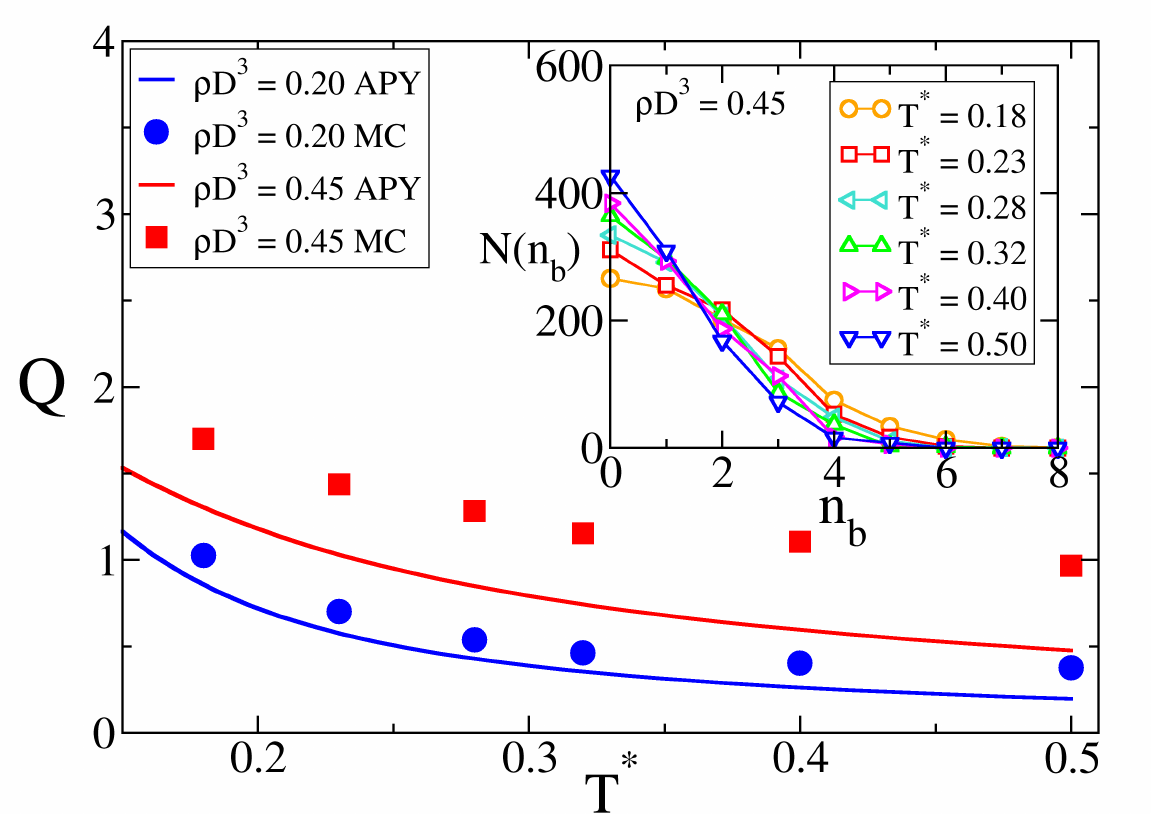}
\caption{(Color online) Average number of bonds per particle $Q$ versus temperature 
  for system M2 at two different densities (as labeled); for
  the definition of $Q$ see text. Symbols correspond to MC data while
  lines represent APY results. The inset displays simulation data for the average number of 
  particles with $n_{\rm b}$ bonds, $N(n_{\rm b})$, versus the number of bonds 
  for selected  temperatures at density $\rho D^3 = 0.45$ (as labeled).}
\label{fig:X-m2}
\end{figure}

The values of $Q$ for model M2 are reported in
Figure \ref{fig:X-m2}.
While the trend in temperature for the agreement between simulation and APY data is similar to model M1,
we observe a significantly different behaviour for the $N(n_b)$-values
as functions of $T$: as a consequence of the larger interaction range,
the number of unbonded particles is now throughout substantially lower
(by about a factor of two), while the number of particles forming more than two bonds is now significantly larger. Probably due to the formation of more than one bond per patch, we observe that the agreement between simulation and theoretical data is not as good as in model M1.

\subsubsection{Thermodynamic properties}

Our analysis of the thermodynamic properties (in terms of the
pressure, $p$, and of the internal energy per particle, $E/N$) is based
on both $NpT$ simulations and theoretical results. Both APY and RHNC data are reported, while BH-TPT data are added only for system M1.

\begin{figure}[h] 
\centering
\includegraphics[width=\columnwidth, clip=true]{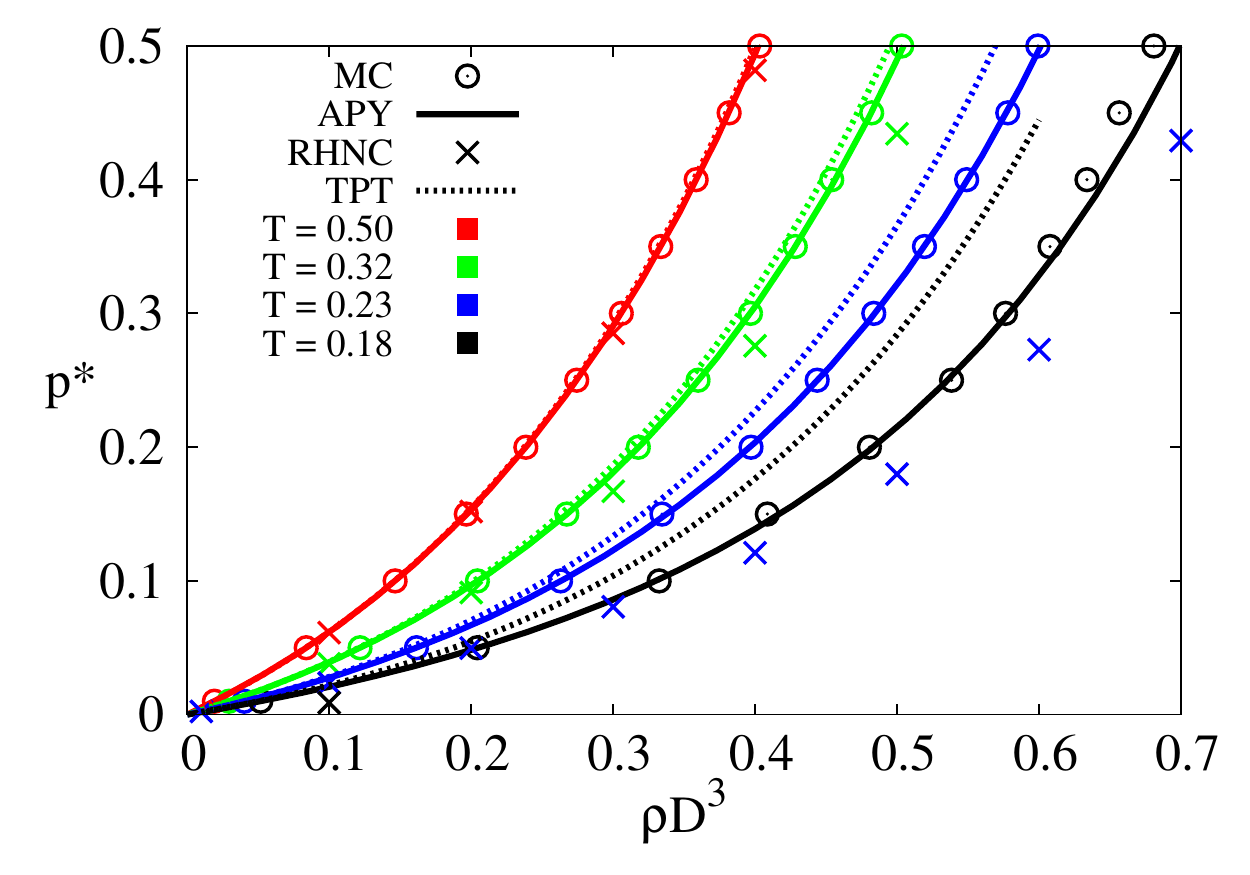}
\includegraphics[width=\columnwidth, clip=true]{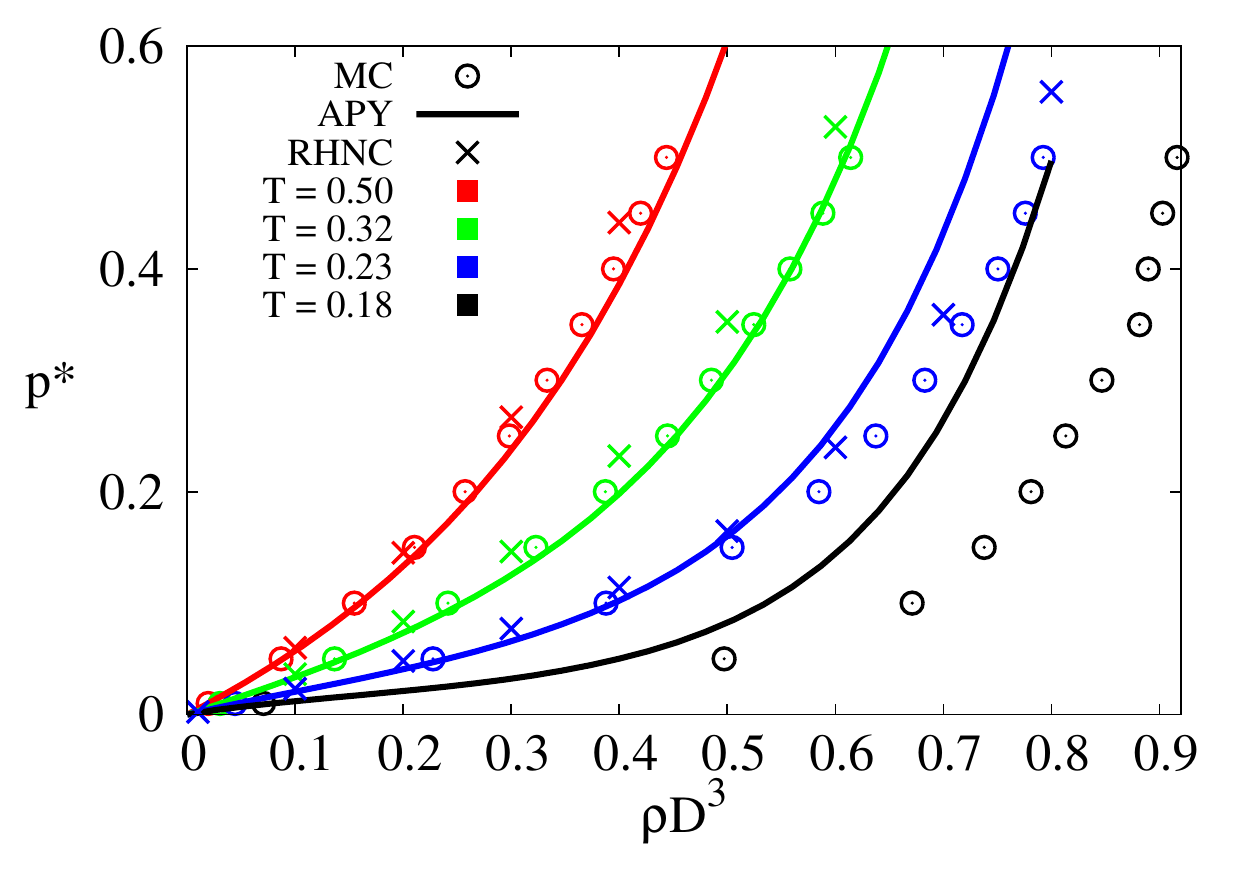}
\caption{(Color online) Pressure $p^*$ versus density for system M1 (top panel) 
  and system M2 (bottom panel) for selected temperatures (as specified by the different colors).
  Open circles -- MC data, continuous lines -- APY data,
  crosses -- RHNC results, and dotted lines -- BH-TPT results (if applicable).}
\label{fig:Prho}
\end{figure}

The pressure has been calculated for our two models along four
isotherms ($T^*=0.5, 0.32, 0.23,$ and 0.18), the corresponding results
are shown in the two panels of Figure \ref{fig:Prho}. 
APY allows to calculate the pressure via the
compressibility [cf. Equation (\ref{Pc})] and via the virial route
[cf. Equation (\ref{Pv})]; since the latter results are considerably
less accurate than the former ones we have not included them in Figure~\ref{fig:Prho}.

Starting with model M1 we see from the top panel of
Figure \ref{fig:Prho} that the results obtained via the different
methods essentially coincide on a single curve for the highest
temperature ($T^* = 0.5$). However, as the temperature decreases this
agreement deteriorates: only the APY-data agree nicely
with the simulation results over the entire temperature range
considered. Further, we observe that -- as compared to the simulation
results -- BH-TPT overestimates these data, while RHNC predicts
systematically smaller values; for the lowest temperature value considered,
the latter approach badly fails for $\rho D^3 \geq 0.1$, leading to negative
pressures that are not showed in the picture.

For the pressure data of model M2 we observe already for the highest temperature
differences between the different approaches which become
more pronounced as the temperature decreases.
These discrepancies are related to the increased bonding volume,
due to the longer interaction range and larger patch angle of model M2.
Also for this model APY seems to be the most reliable theoretical approach,
even though the agreement with MC data is slightly worse than for the previous model.
RHNC works reasonably good, better than for model M1,
but  it does not converge for the lowest temperature;
instead, BH-TPT results turned out to differ by an entire order of magnitude,
showing that this very simple theory is totally inappropriate to
describe systems where the potential differs so substantially from the reference interactions.

\begin{figure}[h] 
\centering
\includegraphics[width=\columnwidth, clip=true]{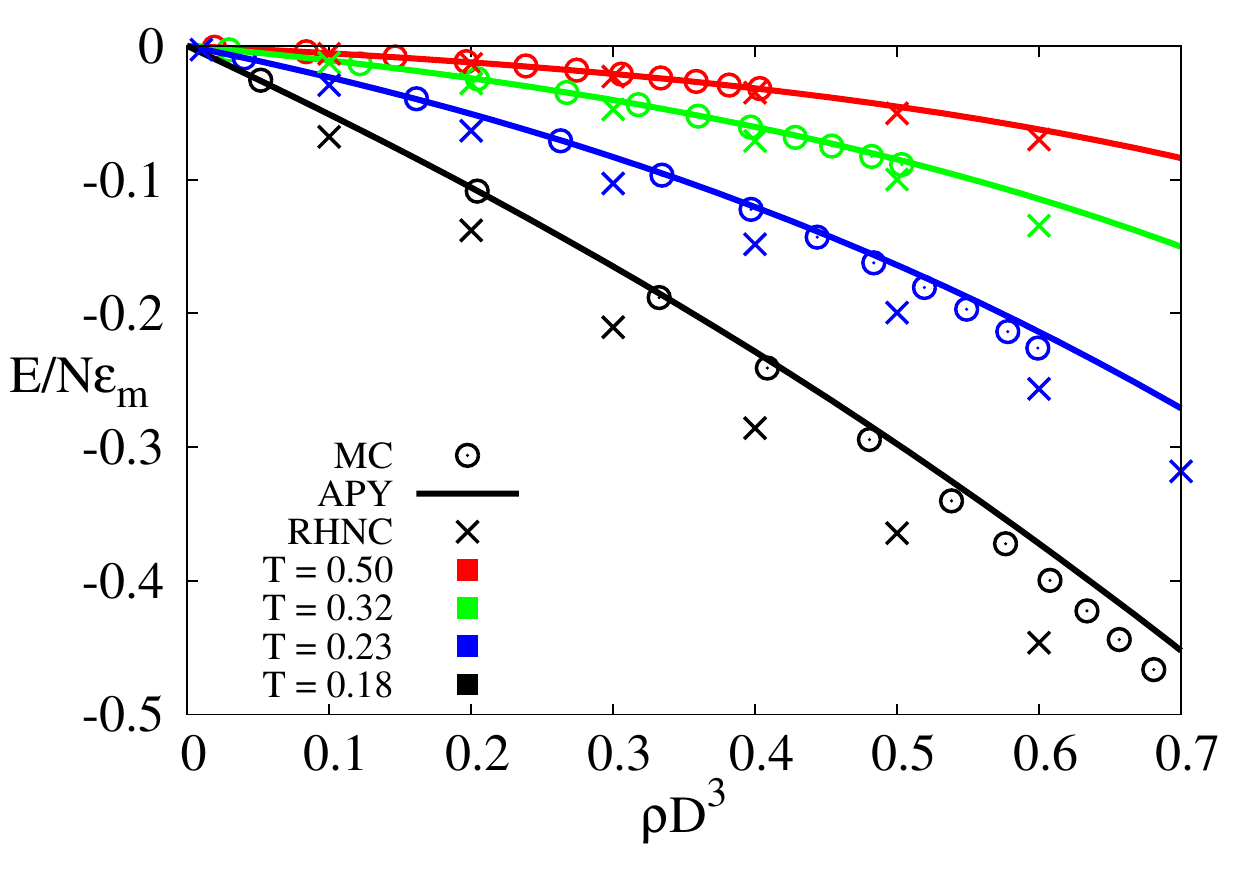}
\includegraphics[width=\columnwidth, clip=true]{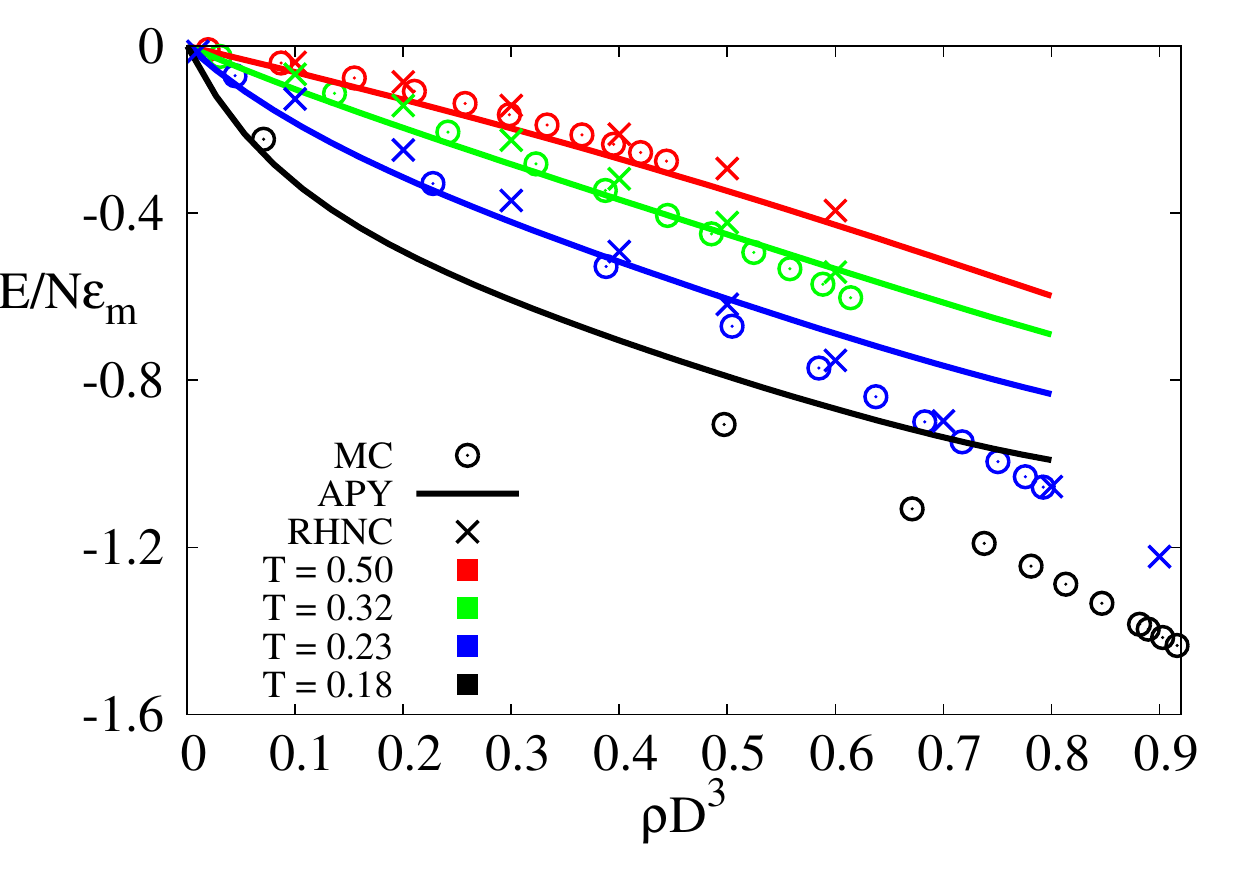}
\caption{(Color online) Internal energy $E/N\epsilon_{\rm m}$ 
  versus density for system M1 (top panel) and system M2 (bottom panel)
  for selected temperatures (as specified by the different colors).
  Open circles -- MC data, continuous lines -- APY data,
  crosses -- RHNC results.}
\label{fig:Erho}
\end{figure}

In the two panels of Figures \ref{fig:Erho} we show the results for
the internal energy per particle $E/N$
for the two models considered as a function of density along the same
four isotherms considered above. Again, good agreement between the
different data sets is observed for model M1 at high and
intermediate temperatures, while differences of up to 15\% occur
along the $(T^* = 0.18$)-isotherm. The agreement for the results obtained for
model M2, shown in the bottom panel of
Figure \ref{fig:Erho}, is less satisfactory. Throughout, APY data are
closer to the simulation results.

\section{Conclusions}
\label{conclusions}
In this contribution we have put forward an extension of the multi-density integral equation formalism proposed by Wertheim~\cite{wertheim1986a,wertheim1986b,wertheim1987,kalyuzhnyi1991} to describe the  properties in the fluid phase of inverse patchy colloids (IPCs), a new class of particles with heterogeneously patterned surfaces. These particles consist of mutually repelling, spherical colloids whose surfaces are decorated by well-defined regions (so-called patches or interaction sites); the patches repel each other while they are attracted by the bare surface of the colloidal particle. Applying Wertheim's formalism, all relevant structural and thermodynamic quantities can be expanded in terms of partial correlation functions, each of them specified by the number of bonds formed by the patches. These functions are obtained from an Ornstein-Zernike (OZ) type integral equation, complemented by closure relations similar to the ones used in standard liquid state theory~\cite{HeM}; the ensuing schemes are called associative liquid state theories. In this contribution we have focused on the associative Percus-Yevick (APY) approach.
The introduction of the ideal network approximation in combination with the orientationally averaged version of the multi-density OZ equation leads to convenient simplifications: in the case of $n_s$ equivalent patches (in our case we have considered two-patch IPCs), only $n_s \times n_s$ partial correlation functions have to be taken into account.
In this associative framework substantially less expansion coefficients have to be considered as compared to standard expansions of structural and thermodynamic quantities in terms of spherical harmonics in systems with directional interactions.

We have applied the associative approach to two different types of two-patch IPCs, that differ substantially in their interaction properties. We have compared the ensuing results with data obtained via Monte Carlo (MC) simulations and via standard liquid state theories, that have been adapted to our model: a standard integral equation based approach -- the reference hypernetted-chain approximation (RHNC) -- and a thermodynamic perturbation theory -- the Barker-Henderson thermodynamic perturbation theory (BH-TPT).
On varying the temperature and the density over a broad range of values, we observe a remarkable agreement between MC data and  both APY and RHNC results for the structural properties; this concerns in particular the contact value and the characteristic shape of the second peak of the pair distribution functions. It worth noting that the APY scheme has a wider range of convergence than RHNC and its agreement with MC data can be followed down to rather low temperatures with the same level of accuracy. Agreement for the thermodynamic data (i.e. for the energy per particle and the pressure at different state points) is satisfactory for high temperatures but deteriorates on lowering the temperature. 

In general, the APY approach proved to be the most stable and reliable theoretical description among the ones considered here. The quality of the APY results was found to be better for IPC systems with relatively small patches and a short interaction range, being these features important in defining the bonding volume of an IPC. In fact, the APY description neglects the possibility of more than one bond per patch.   
We mention that our APY approach has been recently used to describe also the static properties of an IPC system characterized by a short interaction range -- the one used here for the short ranged model -- and a small patch size -- intermediate between the ones of the two models studied in this paper~\cite{silvano30n}. For this system, the APY results proved to be very accurate also in the regime where the dynamics of the system slows down.

The APY theory can also be extended to describe IPC systems with non identical patches. In fact an analtytical as well as a coarse-grained description of IPCs with different patches in either size or charge has been recently proposed~\cite{monika} and could offer an additional application for our associative description.

\section*{Acknowledgments}
The authors would like to thank Fred Lado (Raleigh, USA) for providing the
original RHNC code. E.B., S. F. and G.K. gratefully acknowledge
financial support by the Austrian Science Foundation (FWF) under
Proj. Nos. M1170-N16, V249-N27, P23910-N16, and F41 (SFB ViCoM). The
work was supported within a bilateral ''Wissenschaftlich-Technisches
Abkommen'' project by the \"OAD (Proj. No. UA 04/2013).

\clearpage

\begin{table}
\begin{center}
\begin{tabular}{cccrrrr}
\hline
Model & 
$\;\;\;\;\delta\;\;\;\;\;\;$ & 
$\;\;\;\;\;e\;\;\;\;\;$ &
$\;\;\;\;\;\epsilon_{00}\;\;\;$ & 
$\;\;\;\;\;\;\;\epsilon_{01}\;\;\;\;\;$ & 
$\;\;\;\;\;\;\;\epsilon_{11}\;\;\;\;\;$ &
$\;\;\;\;\;\;\;\epsilon_{\rm m}\;\;\;\;\;$ 
\\
\hline
M1 &  0.1  & 0.3 & 2.8628 & -74.612 & 660.92 & -0.6683\\
M2 &  0.3  & 0.3 & 0.2827 & -6.857 & 57.12   & -0.6683\\
\hline
\end{tabular}
\end{center}
\caption{Model parameters that specify the two different IPC systems
  investigated in this contribution; they are denoted by M1 and
  M2. Values of $\delta$ and $e$ are given in units of the hard-sphere
  diameter, $D$, while energy parameters are given in units of $k_{\rm
    B}T$~\cite{bianchi2011}.}
\label{tab:model-parameters}
\end{table}

\clearpage


\clearpage

\bibliographystyle{unsrt}
\bibliography{biblio-ipc-fluid} 

\end{document}